\documentclass[draft,onecolumn,12pt]{IEEEtran}
\usepackage{enumerate,cite,amsmath,amsfonts,amssymb,subfigure,epsfig,times,graphicx,psfrag}
\usepackage{float}
\usepackage{verbatim,color}
\psfull
\newcommand{\mbf}{\boldmath}
\newcommand{\smbf}{\footnotesize \boldmath}

\def\max{\text{max}}
\def\min{\text{min}}

\def\sbQ{{\mbox {\smbf $Q$}}}
\def\sbI{{\mbox {\smbf $I$}}}
\def\sbH{{\mbox {\smbf $H$}}}

\def\b0{{\mbox {\mbf $0$}}}
\def\bA{{\mbox {\mbf $A$}}}

\def\be{{\mbox {\mbf $e$}}}

\def\bh{{\mbox {\mbf $h$}}}
\def\bp{{\mbox {\mbf $p$}}}

\def\br{{\mbox {\mbf $r$}}}

\def\bu{{\mbox {\mbf $u$}}}

\def\bx{{\mbox {\mbf $x$}}}
\def\bv{{\mbox {\mbf $v$}}}
\def\by{{\mbox {\mbf $y$}}}
\def\bz{{\mbox {\mbf $z$}}}
\def\bD{{\mbox {\mbf $D$}}}

\def\bM{{\mbox {\mbf $M$}}}
\def\bB{{\mbox {\mbf $B$}}}

\def\bH{{\mbox {\mbf $H$}}}
\def\bS{{\mbox {\mbf $S$}}}

\def\sbS{{\mbox {\smbf $S$}}}
\def\sbA{{\mbox {\smbf $A$}}}

\def\bU{{\mbox {\mbf $U$}}}
\def\bI{{\mbox {\mbf $I$}}}
\def\bR{{\mbox {\mbf $R$}}}
\def\bL{{\mbox {\mbf $L$}}}

\def\bQ{{\mbox {\mbf $Q$}}}
\def\bV{{\mbox {\mbf $V$}}}

\def\bb0{{\mathbf{0}}}

\def\b_beta{\mbox{\boldmath $\beta$}}
\def\hb_beta{\mbox{\boldmath $\hat \beta$}}

\newtheorem{theorem}{Proposition}

\newtheorem{Corollary}{Corollary}
\newtheorem{example}{Example}
\newtheorem{definition}{Definition}

\newtheorem{Remark}{Remark}

\begin{document}
\vfill

\title{On Gaussian MIMO BC-MAC Duality With Multiple Transmit Covariance Constraints$^{\star}$\\[0.2cm]}
\author{Lan Zhang$^{\dagger}$, Rui Zhang$^{\ddagger}$, Ying-Chang
Liang$^{\ddagger}$, Yan Xin$^{\dagger\dagger}$, and H. Vincent
Poor$^*$
\thanks{
$^{\star}$This work was supported by the National University of
Singapore (NUS) under start-up grants R-263-000-314-101,
R-263-000-314-112, and R-263-000-478-112, by a NUS Research
Scholarship, and by the U.S. National Science Foundation under
Grants ANI-03-38807 and CNS-06-25637. This work was partially done
while Y. Xin was visiting Princeton University.}
\thanks{$^{\dag}$L. Zhang is with the Department of Electrical and
Computer Engineering, National University of Singapore, Singapore
118622 (email: zhanglan@nus.edu.sg).}
\thanks{$^{\ddag}$R. Zhang and Y.-C. Liang are with Institute of Infocomm Research, A*STAR, 21 Heng Mui Keng
Terrace, Singapore 119613 (email: rzhang@i2r.a-star.edu.sg;
ycliang@i2r.a-star.edu.sg).}
\thanks{$^{\dagger\dagger}$Y. Xin is with the NEC Laboratories America, Princeton, NJ 08540, USA (email: yanxin@nec-labs.com).}
\thanks{$^{*}$H. V. Poor is with the Department of Electrical Engineering, Princeton University, Princeton, NJ 08544, USA (email:
poor@princeton.edu).}} {}
\renewcommand{\thepage}{}
\markboth{}{} \maketitle
\begin{center}
{\vspace{2cm}{Suggested Editorial Areas}}:\\[0.5cm]
Communications, Shannon Theory
\end{center}
\renewcommand{\thepage}{} \maketitle
\newpage
\pagenumbering{arabic} \linespread{1.7}
\begin{abstract}
Owing to the structure of the Gaussian multiple-input
multiple-output (MIMO) broadcast channel (BC), associated
optimization problems such as capacity region computation and
beamforming optimization are typically non-convex, and cannot be
solved directly. One feasible approach to these problems is to
transform them into their dual multiple access channel (MAC)
problems, which are easier to deal with due to their convexity
properties. The conventional BC-MAC duality is established via
BC-MAC signal transformation, and has been successfully applied to
solve beamforming optimization, signal-to-interference-plus-noise
ratio (SINR) balancing, and capacity region computation. However,
this conventional duality approach is applicable only to the case,
in which the base station (BS) of the BC is subject to a single sum
power constraint. An alternative approach is minimax duality,
established  by Yu in the framework of Lagrange duality, which can
be applied to solve the per-antenna power constraint problem. This
paper extends the conventional BC-MAC duality to the general linear
constraint case, and thereby establishes a general BC-MAC duality.
This new duality is applied to solve the capacity computation and
beamforming optimization for the MIMO and multiple-input
single-output (MISO) BC, respectively, with multiple linear
constraints. Moreover, the relationship between this new general
BC-MAC duality and minimax duality is also presented. It is shown
that the general BC-MAC duality offers more flexibility in solving
BC optimization problems relative to minimax duality. Numerical
results are provided to illustrate the effectiveness of the proposed
algorithms.
\end{abstract}

\section{Introduction}\label{sect:intro}

In a Gaussian multiple input multiple output (MIMO) broadcast
channel (BC), the base station (BS) equipped with multiple transmit
antennas sends independent information to each of multiple remote
users, which are equipped with multiple receive antennas. In the
past decade, a great deal of research has been focused on the
characterization of optimal transmission schemes for the MIMO BC
\cite{Liu1998:beamformingpowercontrol,FarrokhiLiu1998,Jindal05:sumpowerMAC,Yuwei2006:sumcapacitycomputation,Yuwei07:perantconst,rzhang06:powerregion}.
Due to the coupled structure of the transmitted signals, the
optimization problems associated with the BC are usually non-convex.
The key technique used to overcome this difficulty is to transform
the BC problem into a convex multiple access channel (MAC) problem
via a so-called BC-MAC duality relationship. Up to now, two types of
BC-MAC duality have been proposed as follows:

\begin{enumerate}
\item {\it Conventional BC-MAC Duality (\cite{Liu1998:beamformingpowercontrol,Jindal05:sumpowerMAC,Tse_02_ISIT:duality,Viswanath2003:sumcapacity}):}\\
      $~~~~$Under a single sum power constraint, the capacity region
      (or signal-to-interference-plus-noise-ratio (SINR) region)
      of a BC is identical to that of a dual MAC under the
      same sum power constraint. The channel matrix associated with the dual MAC
      is the conjugate transposed channel matrix of the BC, and the noise covariance matrices of both channels
      are identity matrices \cite{Liu1998:beamformingpowercontrol,Jindal04:Bstaward}.
\item {\it Minimax Duality (\cite{Yuwei06:dualityminmax,Yuwei07:perantconst,Yuwei2004:broadcastthroughput}):}\\
      $~~~~$The sum rate maximization problem of a BC with multiple linear constraints has the same solution as
      the dual MAC minimax optimization problem. The channel matrix of
      the dual MAC is the conjugate transposed channel matrix of the BC, and the
      noise covariance matrix of the dual MAC is an unknown variable of the minimax optimization
      problem \cite{Yuwei06:dualityminmax}.
\end{enumerate}

The conventional BC-MAC duality was first observed by
Rashid-Rarrokhi {\it et al.} \cite{Liu1998:beamformingpowercontrol},
and applied to solve the sum power minimization problem for a BC
with SINR constraints. Several different methods have been developed
independently to prove the conventional BC-MAC duality. The proof in
\cite{Liu1998:beamformingpowercontrol} is based on the equivalent
transformation that maps the SINR of the MAC to that of the BC.
Vishwanath {\it et al.} \cite{Jindal03:sumcapacity} proved the
conventional BC-MAC duality by presenting the explicit
transformation between the transmit covariance matrix of the BC and
that of the MAC, and applied this duality to solve the sum capacity
problem. Both proofs here are based on the \emph{reciprocity
relation}\cite{Telatar95:CapacityofMIMO} between the BC and its dual
MAC. Another proof based on the Karash-Kuhn-Tucker (KKT) conditions
was given by Visotsky and Madhow \cite{madhow99:duality}. The
conventional BC-MAC duality has been widely applied to many BC
problems. Schubert and Boche \cite{Boche04:sinr_bal} solved an BC
SINR balancing problem, which is to maximize the minimal SINR among
all the users under a sum power constraint, via transforming the
problem into its dual MAC problem. The conventional BC-MAC duality
was also used to show that a dirty paper coding (DPC)
\cite{Costa1983:dirtypapercoding} is a sum-capacity achieving
strategy by Viswanath and Tse \cite{Viswanath2003:sumcapacity}.
Moreover, the entire capacity region for the MIMO BC channel can be
obtained via the conventional BC-MAC duality
\cite{Jindal05:sumpowerMAC,Yuwei2006:sumcapacitycomputation,Shamai06:MIMOBCcapacity}.
However, the conventional BC-MAC duality is applicable only to the
case in which the BS of the BC is subject to a single sum power
constraint.

On the other hand, the sum-capacity for the MIMO-BC was also studied
by Yu and Cioffi \cite{Yuwei2004:broadcastthroughput} via minimax
optimization. The minimax duality was proposed by Yu
\cite{Yuwei06:dualityminmax}, where the conventional BC-MAC duality
and the minimax duality are unified in the framework of Lagrange
duality. However, only the sum capacity is considered in
\cite{Yuwei06:dualityminmax}. Furthermore, Yu and Lan
\cite{Yuwei07:perantconst} extended the minimax duality to solve the
capacity region computation problem and beamforming problem for the
BC with per-antenna power constraint. The proofs of the minimax
duality in \cite{Yuwei07:perantconst} and
\cite{Yuwei06:dualityminmax} are based on Lagrange duality. Compared
with the conventional BC-MAC duality, the minimax duality can be
applied to the case with multiple linear constraints. However, since
the dual MAC problem has a minimax form, and the noise covariance
matrix of the dual MAC is unknown, high-efficiency algorithms, such
as the iterative water-filling algorithm \cite{Yuwei2004:IWF_MAC},
cannot be applied.

\subsection{Overview of the Main Results}

The purpose of this paper is to establish the general BC-MAC duality
via the BC-MAC SINR transformation, and unify the BC-MAC duality and
the minimax duality in the framework of the reciprocity relationship
between the BC and the MAC. By introducing several auxiliary
variables and applying the general BC-MAC duality, the primal BC
problem with \emph{multiple transmit covariance constraints} is
transformed into its dual MAC problem with a single sum power
constraint and can be efficiently solved via the existing algorithm
for its dual MAC as the MAC problem has a convex structure that is
easier to handle.

In this paper, we first consider a MIMO BC with a single general
linear constraint. Relying on the BC-MAC transformation, we prove
that the capacity region of the BC is the same as that of its dual
MAC with a single weighted sum power constraint, which we term {\it
the general BC-MAC duality}. The channel matrix of the dual MAC is
the transposed channel matrix of the primal BC, and its noise
covariance matrix is the coefficient of the linear constraint
instead of being an identity matrix as in the conventional BC-MAC
duality.

To exploit the general BC-MAC duality, the weighted sum rate
maximization problem for the BC with multiple linear constraints is
transformed into a single linear constraint problem by introducing
several auxiliary variables. Though the rate maximization problem
for the BC is a non-convex problem, we show that the KKT conditions
are sufficient for optimality, and show that the subgradient-based
algorithm converges to the optimal solution. A method for obtaining
the subgradient is also given. The relationship between the general
BC-MAC duality based solution and the minimax duality based solution
\cite{Yuwei07:perantconst} is explored. We show that the new method
to handle multiple-constraint is equivalent to that of the minimax
duality. But since the general BC-MAC duality based method solves
the multiple constraint optimization problem in a decoupled manner,
the new result has more flexibility to apply the existing algorithms
for the MAC, while the minimax duality does not. Moreover, we
discuss the weighted sum rate maximization problem with a convex but
nonlinear constraint, and develop a new iterative algorithm to solve
this optimization problem. In addition to the weighted sum rate
maximization problem, the proposed method is also applied to solve
the beamforming problem in a multiple-input single-output (MISO) BC
with multiple linear constraints.

\subsection{Organization and Notation}

The remainder of this paper is organized as follows. The system
model is described in Section \ref{sect:model}. Section
\ref{sect:duality} presents the general BC-MAC duality, where the
transmit covariance matrix of the BC is subject to a linear
constraint. The capacity region computation problem of the BC with
multiple constraints or a single nonlinear constraint is studied in
Section \ref{sect:capacity}. The method to cope with multiple linear
constraints is also applied to solve the beamforming problems in
Section \ref{sect:beamforming}. Several numerical results are
provided in Section \ref{sect:simulation} to illustrate the
effectiveness of the proposed methods. Finally, Section
\ref{sect:conclusions} concludes the paper.

Throughout this paper, we use boldface upper and boldface lower case
letters for matrices and vectors, respectively. $(\cdot)^{H}$
denotes the matrix conjugate transpose operation, and
$\text{tr}(\cdot)$ denotes the matrix trace operation.
$\mathbb{E}(\cdot)$ denotes the expectation operation for random
variables. $\bI$ denotes an identity matrix.

\section{System Model}\label{sect:model}
\begin{figure}
    \begin{minipage}[b]{.48\linewidth}
        \centering
        \vspace{8mm}
        \includegraphics[width = 80mm,height=52mm]{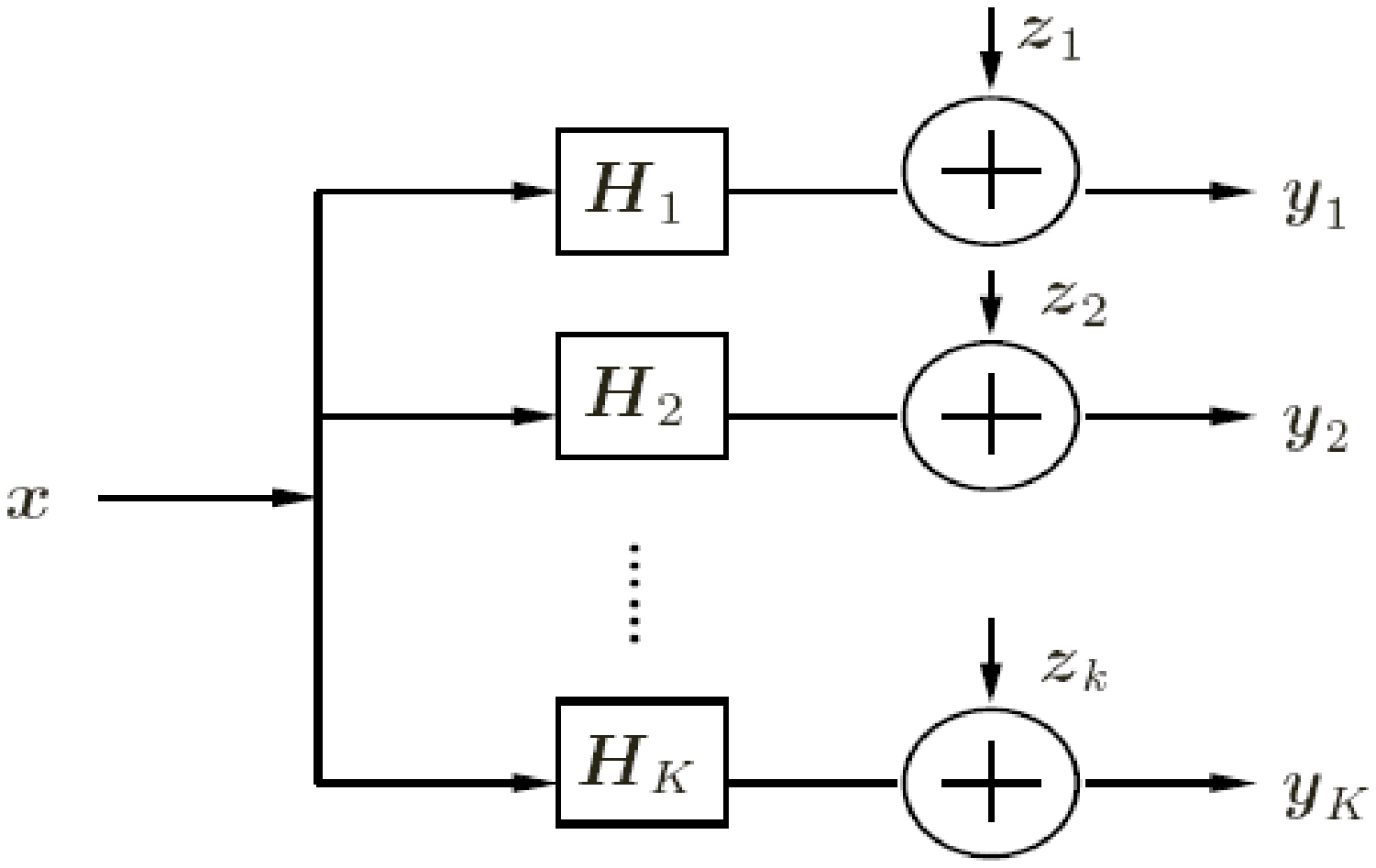}
        \centerline{(a) BC, $\bz_i\sim
        \mathcal{N}(0,\sigma_i^2\bI)$, $\text{\bQ\bA} \leq P$}\medskip
    \end{minipage}
    \hfill
    \begin{minipage}[b]{0.48\linewidth}
        \centering
        \vspace{8mm}
        \includegraphics[width = 80mm,height=46mm]{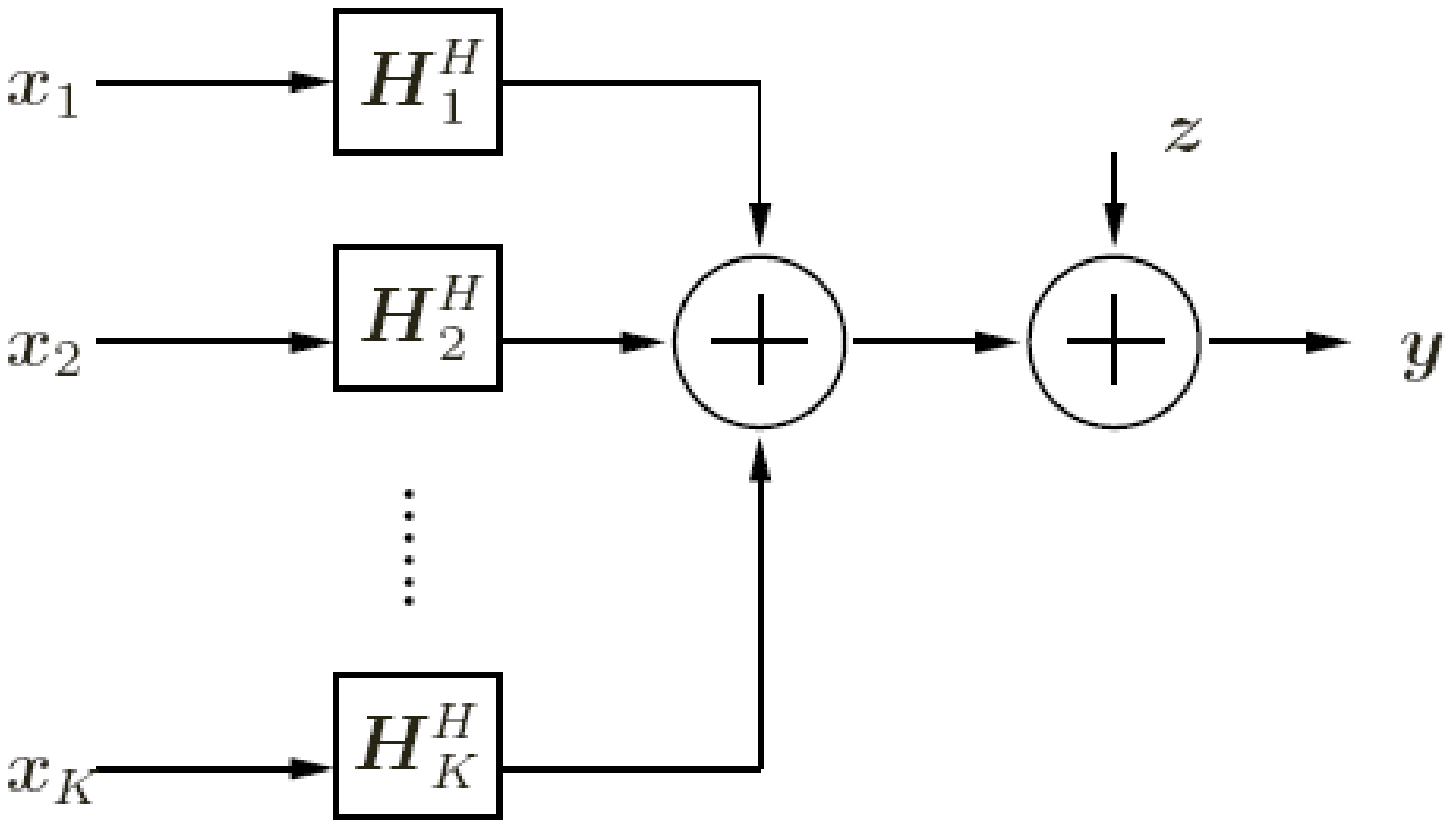}
        \centerline{(b) Dual MAC, $\bz\sim \mathcal{N}(0,\bA)$, $\sum_{i=1}^{K}\sigma_i^2\text{tr}(\bQ_i^{\text{(m)}})\leq P$}\medskip
    \end{minipage}
     \caption{The system models for the primal MIMO BC and the dual MAC.}
     \label{fig:BCMAC}
\end{figure}

We consider a MIMO BC system shown in Fig. \ref{fig:BCMAC} (a),
where the BS intends to send independent information streams to
each of $K$ remote users. The BS has $N_t$ transmit antennas and
each user has $N_r$ receive antennas. The signal received by the
$i$th user is modeled as follows:
\begin{align}\label{def:BC}
\by_i=\bH_i\bx+\bz_i,~i=1,\cdots,K
\end{align}
where the $N_t\times1$ vector $\bx$ denotes the transmit signal at
the BS, $\bH_i$ denotes the $N_r\times N_t$ channel matrix from the
BS to the $i$th user, $\by_i$ denotes the receive signal at the
$i$th user, and $\bz_i$ is the noise vector. The entries of $\bz_i$
are modeled as independent and identically distributed complex
Gaussian random variables with mean zero and variance $\sigma_i^2$.
The transmit signal covariance matrix of the BS is defined as
$\bQ:=\mathbb{E}(\bx\bx^H)$. In this paper, we assume that the
channel knowledge is perfectly known at both the BS and the users,
i.e., $\bH_i$ is perfectly known at the transmitter and the
receivers.

Since the transmit signals for different users are independent, we
have
\begin{align}\label{eq:x}
\bx=\sum_{i=1}^{K}\bx_i
\end{align}
where $\bx_i$ denotes the transmit signal intended for the $i$th
user. Furthermore, to fully utilize the spatial diversity of the
multi-antenna system, a spatial multiplexing scheme is also
applied, which means that the data intended for each user is
further divided into $N$ substreams, where
$N=\min(N_t,N_r)$\cite{Telatar95:CapacityofMIMO}. Thus, the
transmit signal for the $i$th user can be expressed as follows:
\begin{align}\label{eq:xi}
\bx_i=\sum_{j=1}^{N}\bu_{i,j}b_{i,j}
\end{align}
where $b_{i,j}$ is a complex scalar variable with
$p_{i,j}:=\mathbb{E}(|b_{i,j}|^2)$, representing the information
signal of the $j$th data substream of the $i$th user, and
$\bu_{i,j}$ denotes the corresponding beamforming vector with
$||\bu_{i,j}||=1$. Combining \eqref{eq:x}, \eqref{eq:xi} and the
definition of the transmit covariance matrix $\bQ$, we have
\begin{align}\label{eq:Qdecouple}
\bQ=\sum_{i=1}^{K}\sum_{j=1}^{N}p_{i,j}\bu_{i,j}\bu_{i,j}^H.
\end{align}

\subsection{Nonlinear Encoding and Decoding Strategy}
It has been shown that the DPC scheme is a capacity achieving
scheme for the Gaussian MIMO BC \cite{Shamai06:MIMOBCcapacity}.
With the DPC scheme, the information for different users is
encoded in a sequential manner. Without loss of generality, we
assume that the encoding order is identical to the index order,
i.e., the data substream $b_{1,1}$ is encoded first, $b_{1,2}$ is
next encoded, and so on. According to the DPC scheme, the latter
encoded data stream has non-causal information about its former
encoded data streams, and thus the interference caused by the
former data streams' transmission can be completely removed by the
DPC scheme. Thus, the rate achieved by the $i$th user can be
expressed as
\begin{align}
r_i=\log\frac{|\sigma_i^2\sbI+\sbH_i(\sum_{k=i}^{K}\sbQ_i)\sbH_i^H|}{|\sigma_i^2\sbI+\sbH_i(\sum_{k=i+1}^{K}\sbQ_i)\sbH_i^H|}
\end{align}
where $\bQ_i:=\mathbb{E}(\bx_i\bx_i^H)$.

At the receiver side, for each user, successive interference
cancellation (SIC) and the linear minimum mean square error (MMSE)
filter are adopted to decode the corresponding information. With
SIC, the first data stream is decoded by treating all the other
streams as interference; then the signal from the first data stream
is subtracted from the received signal, and the second data stream
is decoded next, and so on. Thus, the mutual information between the
BS and the $i$th user can be expressed as
\begin{align}\label{eq:mutualinfo}
I(\bx_i;\by_i)=I(b_{i,1},\cdots,b_{i,N};\by_i)=I(b_{i,1};\by_i)+I(b_{i,2};\by_i|b_{i,1})+\cdots+I(b_{i,N};\by_i|b_{i,1},\cdots,b_{i,N-1}).
\end{align}
Moreover, since the MMSE receiver is information-lossless
\cite{Tse05:fundwireless}, each term in \eqref{eq:mutualinfo} is
achievable with the MMSE receiver. Thus, the MMSE-SIC receiver can
achieve the capacity of the MIMO system. The receive beamforming
vector for the $j$th data substream at the $i$th user is denoted
by the $N_r\times1$ vector $\bv_{i,j}$. Thus, the SINR of the
$j$th data substream at the $i$th user receiver can be written as
\begin{align}\label{def:SINRbc}
\text{SINR}_{i,j}=\frac{p_{i,j}|\bv_{i,j}^H\bH_i\bu_{i,j}|^2}{\sum_{k=i+1}^{K}\sum_{l=1}^{N}p_{k,l}|\bv_{i,j}^H\bH_i\bu_{k,l}|^2+\sum_{l=j+1}^{N}p_{i,l}|\bv_{i,j}^H\bH_i\bu_{i,l}|^2+\sigma_i^2}.
\end{align}
Since the achievable rate of each data substream depends on its
SINR, we have
\begin{align}\label{eq:rateSINR}
r_i=\sum_{j=1}^{N}\log(1+\text{SINR}_{i,j}).
\end{align}

\subsection{Linear Encoding and Decoding Strategy}

Although the nonlinear strategies DPC and SIC are
capacity-achieving schemes, they are difficult to implement in
practice. A straightforward scheme for transmission is beamforming
without DPC at the transmitter side and SIC at the receiver side.
In the linear strategy, the transmit and receive beamforming
vectors for the $j$th data substream of the $i$th user are still
denoted by $\bu_{i,j}$ and $\bv_{i,j}$, respectively. Thus, the
corresponding SINR can be expressed as
\begin{align}\label{def:sinrlinearbc}
\text{SINR}_{i,j}^{(\text{l})}=\frac{p_{i,j}|\bv_{i,j}^H\bH_i\bu_{i,j}|^2}{\sum_{k=1}^{K}\sum_{l=1}^{N}\
_{(k,l)\neq (i,j)}p_{k,l}|\bv_{i,j}^H\bH_i\bu_{k,l}|^2+\sigma_i^2}.
\end{align}
Since the method developed in the present paper is applicable to
both linear and nonlinear strategies, we mainly focus on the
nonlinear strategy, and adopt the SINR definition \eqref{def:SINRbc}
throughout the paper.

\subsection{General Linear Transmit Covariance Constraint}

In the aforementioned literature, the transmit covariance matrix
is subject only to a sum power constraint or/and a per-antenna
power constraint. In this paper, we consider a general linear
transmit covariance constraint as follows:
\begin{align}\label{eq:cons}
\text{tr}(\bQ\bA)\le P
\end{align}
where $\bA$ is an $N_t\times N_t$ matrix, and $P$ is a constant. If
the matrix $\bA$ is chosen to be an identity matrix, then the
constraint \eqref{eq:cons} is reduced to the sum power constraint;
if the matrix $\bA$ is chosen to be the diagonal matrix having all
diagonal elements being zero except the $j$th element being 1, then
the constraint \eqref{eq:cons} is reduced to the $j$th antenna power
constraint. In cognitive radio networks, we choose $\bA=\bh\bh^H$,
where $\bh$ is the channel response from the secondary transmitter
to the primary receiver, and the constraint is reduced to the
interference power constraint for protecting the primary users
\cite{lan07:jsac,Liang:jstsp,lan:icc08,Lan08TWC:CRBC}. Therefore,
the constraint \eqref{eq:cons} can be viewed as a generalized linear
constraint.

\subsection{Objective Functions}\label{subsect:obj}

In this paper, we consider several scenarios with different
objectives: MIMO BC capacity region computation, SINR balancing,
and power balancing.

$\textit{1) MIMO BC Capacity Region Computation}$

Any boundary rate tuple of the Gaussian MIMO BC capacity region
can be obtained by solving the weighted sum rate maximization
problem with some given user rate weights. Therefore, the capacity
region computation problem is formulated as follows:
\begin{align}
\underset{\sbQ\geq0}{\max}\sum_{i=1}^{K}w_ir_i
\end{align}
where $w_i$ is the positive weight of the $i$th user, and
$\bQ\geq0$ denotes the semidefiniteness constraint. By varying the
values of the weight $w_i$s, the entire capacity region of the
MIMO BC can be obtained.

$\textit{2) SINR Balancing}$

The aim of the SINR balancing problem is to maximize the minimal
ratio between the achieved SINR and the target SINR among all the
data substreams. Mathematically, the optimization problem is
formulated as
\begin{align}\label{obj:balancing}
\underset{\sbQ\geq0}{\max}~\underset{\forall
i,j}{\min}~\frac{\text{SINR}_{i,j}}{\gamma_{i,j}}
\end{align}
where $\gamma_{i,j}$ is the target SINR for the $j$th data
substream of the $i$th user. Conventionally, the SINR balancing
problem considers the MISO case \cite{Boche04:sinr_bal}, i.e.,
$N_t=1$. It has been shown in \cite{Boche04:sinr_bal} that the
ratios of all the data substreams are equal, when the optimal
solution is achieved. Therefore, the problem is termed the SINR
``balancing" problem. Note that \eqref{obj:balancing} is
equivalent to the following form
\begin{align}\label{obj:balancing1}
\begin{split}
\underset{\sbQ\geq0,\alpha}{\max}~&\alpha\\
\text{subject to}~&\text{SINR}_{i,j}\ge \alpha \gamma_{i,j},~\forall
~i,j.
\end{split}
\end{align}
In the sequel, we will use \eqref{obj:balancing1} as the
optimization problem for the SINR balancing problem instead of
\eqref{obj:balancing}, since it is easier to write the Lagrange
function of \eqref{obj:balancing1}.

$\textit{3) Power Balancing}$

In this case, the system has several different power requirements,
and the objective is to minimize the maximal ratio between the
transmit power and the power requirement. Mathematically, the
optimization problem can be expressed as
\begin{align}\label{eq:pwrmin}
\underset{\sbQ\geq0}{\min}~\underset{\forall i}{\max}
~\frac{\text{tr}(\bQ\bA_i)}{P_i}
\end{align}
where $P_i$ is the $i$th power requirement, and $\bA_i$ is an
$N_t\times N_t$ matrix for the $i$th power requirement. Similarly to
the SINR balancing function, the problem \eqref{eq:pwrmin} can be
transformed into
\begin{align}
\begin{split}
\underset{\sbQ\geq0,\alpha}{\min}~&\alpha\\
\text{subject to}~&\text{tr}(\bQ\bA_i)\le \alpha P_i,~\forall i.
\end{split}
\end{align}
If there is a single power requirement, and the corresponding matrix
$\bA$ is an identity matrix, then \eqref{eq:pwrmin} reduces to the
power minimization problem.

In this paper, we will consider these optimization problems with
several general transmit covariance constraints or SINR
constraints.

\section{General BC-MAC Duality}\label{sect:duality}

In this section, we establish the general BC-MAC duality under a
single linear transmit covariance constraint. We start with the SINR
balancing problem expressed as follows:
\begin{align}\label{prob:balancing}
\begin{split}
\underset{\sbQ\geq0,\alpha}{\max}~&\alpha\\
\text{subject to}~&\text{SINR}_{i,j}\ge \alpha
\gamma_{i,j},~\forall ~i,j\\
&\text{tr}(\bQ\bA)\le P.
\end{split}
\end{align}

The MIMO BC SINR balancing problem \eqref{prob:balancing} is a
non-convex optimization problem due to the non-convex SINR
constraints. Although it has been shown in
\cite{Shamai2006:conicoptimazation,Yuwei07:perantconst} that the
SINR constraint under the MISO scenario can be transformed into the
second order cone (SOC) form, the transformation is not applicable
to the MIMO case due to the essentially non-convex property of the
MIMO SINR constraints. Hence, the problem \eqref{prob:balancing} is
still an open problem, and cannot be solved via existing methods.
However, we can establish a new MAC called the dual MAC, and
formulate a dual MAC problem of the primal problem
\eqref{prob:balancing} such that it shares the same solution as its
primal problem \eqref{prob:balancing}.
\begin{definition}
The dual MAC of the primal BC in \eqref{def:BC} has the conjugate
transposed channel matrix of the BC, i.e., the channel matrix of the
dual MAC from the $i$th user to the BS is $\bH_i^H$, and the noise
covariance matrix at the BS is the matrix $\bA$ instead of the
identity matrix, which is shown in the Fig. \ref{fig:BCMAC} (b).
\end{definition}

The corresponding dual MAC optimization problem is expressed as
follows:
\begin{align}\label{prob:dualmacbal}
\begin{split}
\underset{\sbQ^{(m)}_i\geq0,~\alpha}{\max}~&\alpha\\
\text{subject to}~&\text{SINR}_{i,j}^{(m)}\ge \alpha
\gamma_{i,j},~\forall ~i,j\\
&\sum_{i=1}^{K}\sigma_i^2\text{tr}(\bQ^{(m)}_i)\le P
\end{split}
\end{align}
where $\bQ^{(m)}_i$ denotes the transmit signal covariance matrix of
the $i$th user, $\text{SINR}_{i,j}^{(m)}$ denotes the SINR of the
$j$th data substream of the $i$th user, and the superscript
`${(m)}$' indicates that the corresponding variables are for the
dual MAC. In this dual MAC, the MMSE-SIC scheme is applied, which
means that the data streams of the dual MAC are decoded in a
sequential manner. In the dual MAC problem \eqref{prob:dualmacbal},
the decoding order at the BS is converse to the encoding order of
the DPC at the primal BC. Accordingly, the $\text{SINR}_{i,j}^{(m)}$
can be expressed as
\begin{align}\label{def:sinrmac}
\text{SINR}_{i,j}^{(m)}=\frac{q_{i,j}|\bu_{i,j}\bH_i^H\bv_{i,j}|^2}{\bu_{i,j}^H\Big(\sum_{k=1}^{i-1}\sum_{l=1}^{N}q_{k,l}\bH_k^H\bv_{k,l}\bv_{k,l}^H\bH_k+\sum_{l=1}^{j-1}q_{i,l}\bH_i^H\bv_{i,l}\bv_{i,l}^H\bH_i+\bA\Big)\bu_{i,j}}
\end{align}
where $q_{i,j}$ is the power allocated to this data substream, and
$\bu_{i,j}$ and $\bv_{i,j}$ denote the corresponding receive and
transmit beamforming vectors, respectively. While it may be
somewhat confusing at first that the beamforming vectors of the
dual MAC share the same notation with the beamforming vectors of
the primal BC in \eqref{def:SINRbc}, it will become clear in the
following that the optimal beamforming vectors of the primal BC
are identical to those of the dual MAC.

The relationship between the primal problem \eqref{prob:balancing}
and the dual MAC problem \eqref{prob:dualmacbal} is summarized in
the following proposition.
\begin{theorem}\label{thm:SINRduality}
The optimal solutions of the primal problem \eqref{prob:balancing}
and the dual MAC problem \eqref{prob:dualmacbal} are the same.
\end{theorem}

\begin{proof}
First, we will prove by contradiction that the SINR constraints for
problem \eqref{prob:balancing} hold with equality when the optimal
solutions are achieved. If $\text{SINR}_{i,j}>\alpha^*\gamma_{i,j}$
with $\alpha^*$ being the optimal solution of problem
\eqref{prob:balancing}, then we can reduce part of the power
$p_{i,j}$ and distribute it to all the other data substreams,
thereby increasing the objective value $\alpha$ without violating
the constraints. This contradicts $\alpha^*$ being the optimal
solution. A similar argument holds for the problem
\eqref{prob:dualmacbal}.

We next prove that if $\bar{\alpha}$ is achievable for the problem
\eqref{prob:balancing}, then it is also achievable for the problem
\eqref{prob:dualmacbal}. Assume that when $\bar{\alpha}$ is
achieved, $\bar{\bu}_{i,j}$ and $\bar{\bv}_{i,j}$ are the
corresponding beamforming vectors for transmitter and receiver,
respectively, and $\bar{p}_{i,j}$ is the power allocated to the
$j$th data substream of the $i$th user. For the dual MAC problem
\eqref{prob:dualmacbal}, we can choose $\bar{\bv}_{i,j}$ to be the
transmit beamforming vector for the user side, and $\bar{\bu}_{i,j}$
to be the receive beamforming vector at the BS. The power allocated
to the $j$th data substream of the $i$th user of the dual MAC is
assumed to be $\bar{q}_{i,j}$, which can be obtained by setting
$\text{SINR}^{(m)}_{i,j}=\text{SINR}_{i,j}=\bar{\alpha}\gamma_{i,j}$,
i.e.,
\begin{align}\label{eq:sinrequal}
\bar{\alpha}\gamma_{i,j}=&\frac{\bar{p}_{i,j}|\bar{\bv}_{i,j}^H\bH_i\bar{\bu}_{i,j}|^2}{\sum_{k=i+1}^{K}\sum_{l=1}^{N}\bar{p}_{k,l}|\bar{\bv}_{i,j}^H\bH_i\bar{\bu}_{k,l}|^2+\sum_{l=j+1}^{N}\bar{p}_{i,l}|\bar{\bv}_{i,j}^H\bH_i\bar{\bu}_{i,l}|^2+\sigma_i^2}\notag\\
=&\frac{\bar{q}_{i,j}|\bar{\bu}_{i,j}\bH_i^H\bar{\bv}_{i,j}|^2}{\bar{\bu}_{i,j}^H\Big(\sum_{k=1}^{i-1}\sum_{l=1}^{N}\bar{q}_{k,l}\bH_k^H\bar{\bv}_{k,l}\bar{\bv}_{k,l}^H\bH_k+\sum_{l=1}^{j-1}\bar{q}_{i,l}\bH_i^H\bar{\bv}_{i,l}\bar{\bv}_{i,l}^H\bH_i+\bA\Big)\bar{\bu}_{i,j}}.
\end{align}
By rearranging \eqref{eq:sinrequal}, we can list all the equations
related to the SINR as follows:
\begin{align}
&\bar{p}_{1,1}\Big(\bar{\bu}_{1,1}^H\bA\bar{\bu}_{1,1}\Big)=\bar{q}_{1,1}\Big(\sum_{k=2}^{K}\sum_{l=1}^{N}\bar{p}_{k,l}|\bar{\bv}_{1,1}^H\bH_1\bar{\bu}_{k,l}|^2+\sum_{l=2}^{N}\bar{p}_{1,l}|\bar{\bv}_{1,1}^H\bH_1\bar{\bu}_{1,l}|^2+\sigma_1^2\Big)\notag\\
&\bar{p}_{1,2}\Big(\bar{\bu}_{1,2}^H\big(\bar{q}_{1,l}\bH_1^H\bar{\bv}_{1,l}\bar{\bv}_{1,l}^H\bH_1+\bA\big)\bar{\bu}_{1,2}\Big)=\bar{q}_{1,2}\Big(\sum_{k=2}^{K}\sum_{l=1}^{N}\bar{p}_{k,l}|\bar{\bv}_{1,2}^H\bH_1\bar{\bu}_{k,l}|^2+\!\!\sum_{l=3}^{N}\bar{p}_{1,l}|\bar{\bv}_{1,2}^H\bH_1\bar{\bu}_{1,l}|^2\!+\!\sigma_1^2\Big)\notag\\
&\ \ \ \ \ \ \ \ \ \ \ \ \vdots\hspace*{6cm}\vdots\notag\\
&\bar{p}_{K,N}\Big(\bar{\bu}_{K,N}^H\Big(\sum_{k=1}^{K-1}\sum_{l=1}^{N}\bar{q}_{k,l}\bH_k^H\bar{\bv}_{k,l}\bar{\bv}_{k,l}^H\bH_k+\sum_{l=1}^{N-1}\bar{q}_{K,l}\bH_K^H\bar{\bv}_{K,l}\bar{\bv}_{K,l}^H\bH_K+\bA\Big)\bar{\bu}_{K,N}\Big)=\bar{q}_{K,N}\sigma_K^2\notag.
\end{align}
By adding the above equations together, we have
\begin{align}\label{eq:powequal}
\sum_{i=1}^{K}\sum_{j=1}^{N}\bar{p}_{i,j}\bar{\bu}_{i,j}^H\bA\bar{\bu}_{i,j}=\sum_{i=1}^{K}\sum_{j=1}^{N}\sigma_i^2\bar{q}_{i,j}.
\end{align}

From the power constraint $\text{tr}(\bQ\bA)\le P$ of the problem
\eqref{prob:balancing}, we have
$\sum_{i=1}^{K}\sum_{j=1}^{N}\bar{p}_{i,j}\bar{\bu}_{i,j}^H\bA\bar{\bu}_{i,j}\le
P$. Combining this with \eqref{eq:powequal}, we have
$\sum_{i=1}^{K}\sum_{j=1}^{N}\sigma_i^2\bar{q}_{i,j}\le P$, which
means that the weighted sum power constraint
$\sum_{i=1}^K\sigma_i^2\text{tr}(\bQ_i^{(m)})\le P$ of the problem
\eqref{prob:dualmacbal} is satisfied. Hence, $\bar{\alpha}$ is
achievable for the problem \eqref{prob:dualmacbal}.

Contrarily, we can prove that if $\tilde{\alpha}$ is achievable for
the problem \eqref{prob:dualmacbal}, then $\tilde{\alpha}$ is also
achievable for the problem \eqref{prob:balancing}.

The proof of Proposition \ref{thm:SINRduality} follows.
\end{proof}

If we assume that the optimal solution for both problems
\eqref{prob:balancing} and \eqref{prob:dualmacbal} is $\alpha^*$,
then it means that the point
$\{\text{SINR}_{i,j}=\alpha^*\gamma_{i,j},\forall~i,j\}$ is
achievable for the primal BC and the dual MAC with corresponding
constraints, respectively. Therefore, under the general linear
constraint $\text{tr}(\bQ\bA)\le P$, the primal BC can achieve the
same SINR region as the dual MAC, which is subject to a weighted
sum power constraint. From an information theoretic perspective,
according to \eqref{eq:rateSINR}, the rate point
$\{\br=[r_1,\cdots,r_K]\}$ is achievable if the SINR point
$\{\text{SINR}_{i,j},~\forall i,j\}$ is achievable under
corresponding constraints. Thus, we have the following corollary.
A rigorous proof is provided in Appendix \ref{prov:cor}.

\begin{Corollary}\label{cor:rateregion}
The capacity region of the primal BC under the constraint
$\text{tr}(\bQ\bA)\le P$, is equal to the capacity region of its
dual MAC with a single weighted sum power constraint
$\sum_{i=1}^{K}\sigma_i^2\text{tr}(\bQ^{(m)}_i)\le P$.
\end{Corollary}

\begin{Remark}
By setting $\bA$ in \eqref{prob:balancing} to be an identity
matrix, and assuming that $\sigma_i^2=1$ for all users, the
general linear power constraint becomes a sum power constraint,
and the noise covariance at the BS of the dual MAC is reduced to
an identity matrix. This is precisely the same as the conventional
BC-MAC duality. Thus, the new BC-MAC duality can be viewed as a
generalization of the conventional BC-MAC duality. The proof of
the duality is based on the special BC-MAC reciprocity
relationship, instead of the Lagrange duality used in
\cite{Yuwei06:dualityminmax,Yuwei07:perantconst}. Note that since
the SINR constraints in \eqref{prob:balancing} is not convex, the
Lagrange duality gap between \eqref{prob:balancing} and
\eqref{prob:dualmacbal} may not be zero. Therefore, Proposition
\ref{thm:SINRduality} cannot be proved through the use of Lagrange
duality. But for the capacity region problem, the objective
function is concave in the signal covariance matrices and convex
in the noise covariance matrices. Hence, the Lagrange duality gap
is zero and the Lagrange duality can be applied for the proof of
the BC-MAC capacity duality. From this perspective, the
reciprocity relationship is more fundamental than the Lagrange
duality for the BC-MAC duality.
\end{Remark}

\begin{Remark}
For a given set of transmit covariance matrices of the dual MAC, we
can obtain the corresponding transmit covariance matrix of the
primal BC to achieve the same value of $\alpha$ by using the method
giving in the proof of Proposition \ref{thm:SINRduality}. The
detailed MAC-BC covariance matrix transformation algorithm is
provided in Table \ref{tabl:transformSINR}. Similarly, the BC-MAC
covariance matrix transformation algorithm is readily obtained.
Furthermore, the proof of Corollary \ref{cor:rateregion} presents a
MAC-BC covariance matrix transformation such that the primal BC and
its dual MAC achieve the same rate tuple. The detailed algorithm is
presented in Table \ref{tabl:transform}.
\end{Remark}

\begin{table}
\caption{\label{tabl:transformSINR} The MAC-BC covariance
transformation for SINR equivalence.}
\begin{center}
\begin{tabular}{l}
  \hline
  MAC-BC covariance matrix transformation I\\
  \hline
   1. Apply eigenvalue decomposition to $\bQ_i^{(m)}=\bV_i\mathbf{\Lambda}_i\bV_i$ with $\bv_{i,j}$ being the $j$th column of $\bV_i$ \\
   $~~~~~~~~$and $q_{i,j}$ being the $j$th diagonal element of
   $\mathbf{\Lambda}_i$,\\
   2. For each data substream, apply the MMSE algorithm to compute the receive beamforming vector $\bu_{i,j}$, i.e., \\
   $~~~~~~~~$$\bu_{i,j}=\Big(\sum_{k=1}^{i-1}\sum_{l=1}^{N}\bar{q}_{k,l}\bH_k^H\bar{\bv}_{k,l}\bar{\bv}_{k,l}^H\bH_k+\sum_{l=1}^{j-1}\bar{q}_{i,l}\bH_i^H\bar{\bv}_{i,l}\bar{\bv}_{i,l}^H\bH_i+\bA\Big)^{-1}\bH_i^H\bar{\bv}_{i,j}$.\\
   3. According to \eqref{eq:sinrequal}, compute $\bp_{i,j}$, $\forall~i,j$, \\
   4. $\bQ_i=\sum_{j=1}^{N}p_{i,j}\bu_{i,j}\bu_{i,j}^H$. \\
  \hline
\end{tabular}
\end{center}
\end{table}

\begin{table}
\caption{The MAC-BC covariance transformation algorithm for capacity
equivalence.}
\begin{center}
\begin{tabular}{l}
  \hline
  MAC-BC covariance matrix transformation II \\
  \hline
   1. Define $\hat{\bH}_k = \bH_k\bA^{-1/2}$, $\bM_k = \Big(\bI+\sum_{j=i+1}^{K}\hat{\bH}_j^H\bQ_j^{(m)}\hat{\bH}_j\Big)$ and  \\
   $~~~~~~~~$ $\bB_k =
   \Big(\bI+\hat{\bH}_k\big(\sum_{j=i+1}^{K}\bQ_j\big)\hat{\bH}_k^H\Big)$\\
   2. $\mathbf{for}$ $k=1,\cdots,K$ \\
   3. $~~~~~$calculate the SVD decomposition of $\bM_k^{-1/2}\hat{\bH}_k\bB_k^{-1/2}=\bR_k\bD_k\bL_k$\\
   4. $~~~~~$$\bQ_k=\bA^{-1/2}\bM_k^{-1/2}\bR_k\bL_k^H\bB_k^{1/2}\bQ_k^{(m)}\bB_k^{1/2}\bL_k\bR_k^H\bM_k^{-1/2}(\bA^{-1/2})^H$\\
  5. $\mathbf{end~for}$\\
  \hline
\end{tabular}\label{tabl:transform}
\end{center}
\end{table}

Moreover, although Proposition \ref{thm:SINRduality} is for the
nonlinear scheme, it is also applicable to the linear transmission
scheme.
\begin{Corollary}
Under the linear transmit strategy, the achievable SINR region of
the primal BC under the constraint $\text{tr}(\bQ\bA)\le P$, is
equal to the achievable SINR region of its dual MAC with a single
weighted sum power constraint
$\sum_{i=1}^{K}\sigma_i^2\text{tr}(\bQ^{(m)}_i)\le P$.
\end{Corollary}

In the following, we will show how to use this general BC-MAC
duality to solve various BC optimization problems with multiple
transmit covariance constraints.

\section{Capacity Computation}\label{sect:capacity}

In the preceding section, the duality between the MIMO BC and the
dual MAC with a general linear constraint was presented. In this
section, by exploiting this duality, we will compute the capacity
region of the MIMO BC. According to the discussion in Section
\ref{subsect:obj}, the capacity of the BC can be obtained by solving
the weighted sum rate maximization problem. For simplicity, we
assume in the sequel that $\sigma_i^2=1$ for all the users.

\subsection{Single Linear Transmit Covariance Constraint}

We first consider the weighted sum rate maximization problem for the
BC with a single linear constraint, which is formulated as follows:
\begin{align}\label{prob:wr1con}
\begin{split}
\underset{\sbQ\geq0}{\max}~&\sum_{i=1}^{K}w_ir_i\\
\text{subject to}~&\text{tr}(\bQ\bA)\le P
\end{split}
\end{align}
where $\bA$ is a constant matrix, and $P$ is a constant. The
problem \eqref{prob:wr1con} is a non-convex problem, and thus
cannot be solved directly. According to Corollary
\ref{cor:rateregion}, the problem \eqref{prob:wr1con} is
equivalent to its dual MAC problem as follows:
\begin{align}\label{prob:dualwr1}
\begin{split}
\underset{\sbQ_i^{(m)}\geq0}{\max}~&\sum_{i=1}^{K}w_ir_i^{(m)}\\
\text{subject to}~&\sum_{i=1}^{K}\text{tr}(\bQ_i^{(m)})\le P
\end{split}
\end{align}
where
$r_i^{(m)}:=\log\frac{|\sbA+\sum_{k=1}^{i}\sbH_k^H\sbQ_{k}^{(m)}\sbH_k|}{|\sbA+\sum_{k=1}^{i-1}\sbH_k^H\sbQ_{k}^{(m)}\sbH_k|}$
denotes the achievable rate of the $i$th user. By solving the
problem \eqref{prob:dualwr1} via the interior point algorithm
\cite{Boyd_optimization_book}, the optimal solution for the problem
\eqref{prob:wr1con} can be obtained via a MAC-BC transmit covariance
matrix transformation algorithm.

In the following, we present an important property of the problem
\eqref{prob:wr1con}, which will be used in the case with multiple
transmit covariance constraints. We first list the KKT conditions of
the problem \eqref{prob:wr1con} as follows:
\begin{align}\label{eq:kktwr1}
\begin{split}
\frac{\partial\sum_{i=1}^{K}w_ir_i}{\partial
\sbQ_i}=\lambda\bA+\mathbf{\Psi}_i,~\forall i\\
\lambda\Big(\text{tr}(\bQ\bA)-P\Big)=0
\end{split}
\end{align}
where $\lambda$ is the Lagrange multiplier, and $\mathbf{\Psi}_i$ is
the Lagrange multiplier associated with the constraint $\bQ_i\geq0$.
In general, the KKT conditions are only necessary for a solution to
be optimal for a non-convex problem. However, for the problem
\eqref{prob:wr1con}, it is shown in the following proposition that
the KKT conditions are also sufficient for optimality.

\begin{theorem}\label{thm:kktsuf}
The KKT conditions \eqref{eq:kktwr1} are sufficient for a solution
to be optimal for the problem \eqref{prob:wr1con}.
\end{theorem}
\begin{proof}
According to the Corollary \ref{cor:rateregion}, the problem
\eqref{prob:wr1con} is equivalent to its dual MAC problem
\eqref{eq:kktwr1}. We now assume that $\tilde{\bQ}$ satisfies the
KKT conditions in \eqref{eq:kktwr1} and achieves the weighted sum
rate $\tilde{R}$. Then, by the BC-MAC transmit covariance matrix
transformation, we can obtain a set of $\tilde{\bQ}_{i}^{(m)}$s
for the problem \eqref{prob:dualwr1} to achieve the same
$\tilde{R}$. We next assume that $\bar{\bQ}$ is an optimal
solution of the problem \eqref{prob:wr1con} with the optimal
weighted sum rate $\bar{R}$, where $\bar{R}>\tilde{R}$. Thus, we
can obtain the optimal solution of the problem
\eqref{prob:dualwr1} $\bar{\bQ}_i^{(m)}$ by MAC-BC transmit
covariance matrix transformation.

It is well known that the objective function of \eqref{prob:dualwr1}
is a convex function. Hence, we have
$\bQ_i^*:=\tilde{\bQ}_{i}^{(m)}+t\Big(\bar{\bQ}_i^{(m)}-\tilde{\bQ}_i^{(m)}\Big)$,
where $0<t<1$, is a better solution than $\tilde{\bQ}_i^{(m)}$ for
the problem \eqref{prob:dualwr1}. Through the MAC-BC transmit
covariance matrix transformation, we transform the dual MAC solution
$\bQ_i^*$ into its corresponding BC solution $\bQ^*$. Since the
MAC-BC transmit covariance matrix transformation is continuous, we
can always find a $t$ such that $\|\tilde{\bQ}-\bQ^*\|\le \epsilon$
for a given $\epsilon>0$. Therefore, $\tilde{\bQ}$ is not the local
optimal solution, which is contradicted with the KKT condition
$\partial(\sum_{i=1}^{K}w_ir_i)/{\partial
\sbQ_i}=\lambda\bA+\mathbf{\Psi}_i$. The proof thus follows.
\end{proof}

\subsection{Multiple Linear Transmit Covariance Constraints}\label{subsect:mlinearconcap}

We now consider the weighted sum rate maximization problem with
multiple linear constraints as follows:
\begin{align}\label{prob:capcomp}
\begin{split}
\underset{\sbQ\geq0}{\max}~&\sum_{i=1}^{K}w_ir_i\\
\text{subject to}~&\text{tr}(\bQ\bA_1)\le P_1\\
&\text{tr}(\bQ\bA_2)\le P_2
\end{split}
\end{align}
where $\bA_i$, $i=1,2$, is an $N_t\times N_t$ constant matrix, and
$P_i$, $i=1,2$, is a constant. For convenience of description, we
discuss only the case of two linear constraints, though our method
can be easily extended to the case of an arbitrary number of
linear constraints.

Since the objective function is non-concave in $\bQ$, the problem
\eqref{prob:capcomp} is not convex, and thus cannot be solved
directly. In
\cite{Jindal05:sumpowerMAC,Yuwei2006:sumcapacitycomputation}, the
sum capacity of the MIMO BC with a single sum-power constraint was
studied. Based on the conventional BC-MAC duality
\cite{Jindal03:sumcapacity}, the BC problem was transformed into
its dual convex MAC problem with a single sum-power constraint.
However, the problem \eqref{prob:capcomp} is with multiple
constraints, and thus the conventional BC-MAC duality cannot be
applied. To exploit the general BC-MAC duality, we can transform
the problem \eqref{prob:capcomp} into the following problem with a
single constraint:
\begin{align}\label{prob:equalcap}
\begin{split}
g(\lambda_1,\lambda_2):=&\underset{\sbQ\geq0}{\max}~\sum_{i=1}^{K}w_ir_i\\
\text{subject
to}~\lambda_1\text{tr}(\bQ\bA_1)&+\lambda_2\text{tr}(\bQ\bA_2)\le
\lambda_1P_1+\lambda_2P_2
\end{split}
\end{align}
where $\lambda_1$ and $\lambda_2$ can be viewed as auxiliary
variables. The relationship between the problem \eqref{prob:capcomp}
and the problem \eqref{prob:equalcap} can be summarized as follows.
\begin{theorem}\label{thm:upperbnd}
The optimal solution of the problem \eqref{prob:equalcap} is an
upper bound on that of the problem \eqref{prob:capcomp}.
\end{theorem}
\begin{proof}
If $\bQ$ is feasible for the problem \eqref{prob:capcomp}, then it
is also feasible for the problem \eqref{prob:equalcap}. Therefore,
the feasible region of the problem \eqref{prob:capcomp} is a subset
of that of the problem \eqref{prob:equalcap}. The proof follows
immediately.
\end{proof}

Furthermore, we can prove that the upper bound is tight, i.e., the
optimal solution of the problem \eqref{prob:capcomp} achieves the
upper bound $g(\lambda_1,\lambda_2)$ for some $\lambda_1$ and
$\lambda_2$. Thus, we have
\begin{theorem}\label{thm:upbndachiev}
The optimal solution of the problem \eqref{prob:capcomp} is equal to
that of the problem
$\min_{\lambda_1,\lambda_2}g(\lambda_1,\lambda_2)$.
\end{theorem}
\begin{proof}
The KKT condition of the problem \eqref{prob:capcomp} can be listed
as follows:
\begin{align}
\begin{split}
\frac{\partial\sum_{i=1}^{K}w_ir_i}{\partial
\sbQ_i}=\mu_1\bA_1+\mu_2\bA_2+\mathbf{\Omega}_i,~\forall i\\
\mu_1\Big(\text{tr}(\bQ\bA_1)-P_1\Big)=0\\
\mu_2\Big(\text{tr}(\bQ\bA_2)-P_2\Big)=0
\end{split}
\end{align}
where $\mu_1$ and $\mu_2$ are the Lagrange multipliers with respect
to the two constraints, respectively, and $\mathbf{\Omega}_i$ is the
Lagrange multiplier associated with the constraint $\bQ_i\geq0$.
When the optimal solution of the problem \eqref{prob:capcomp} is
achieved, we assume that the corresponding optimal variables are
$\bQ^*$, $\mu_1^*$, $\mu_2^*$, and $\mathbf{\Omega}_i^*$.

We now list the KKT conditions of the problem \eqref{prob:equalcap}
as follows:
\begin{align}\label{kkt:equalcap}
\begin{split}
\frac{\partial\sum_{i=1}^{K}w_ir_i}{\partial
\sbQ_i}=\lambda(\lambda_1\bA_1+\lambda_2\bA_2)+\mathbf{\Upsilon}_i,~\forall i\\
\lambda\Big(\lambda_1\text{tr}(\bQ\bA_1)+\lambda_2\text{tr}(\bQ\bA_2)-\mu_1P_1-\mu_2P_2\Big)=0
\end{split}
\end{align}
where $\lambda$ is the Lagrange multiplier, and
$\mathbf{\Upsilon}_i$ is the Lagrange multiplier associated with the
constraint $\bQ_i\geq0$. If we choose $\bQ=\bQ^*$, $\lambda=1$,
$\lambda_1=\mu_1^*$, $\lambda_2=\mu_2^*$, and
$\mathbf{\Upsilon}_i=\mathbf{\Omega}_i^*$, the KKT conditions
\eqref{kkt:equalcap} are satisfied. Since the problem
\eqref{prob:equalcap} is the weighted sum rate maximization problem
with a single linear constraint, according to Proposition
\ref{thm:kktsuf}, the solution is the optimal solution of the
problem \eqref{prob:equalcap}. Combining this with Proposition
\ref{thm:upperbnd}, the proof follows.
\end{proof}

According to the general BC-MAC duality discussed in Section
\ref{sect:duality}, the problem \eqref{prob:equalcap} is
equivalent to the following dual MAC problem:
\begin{align}\label{prob:innercapdual}
\begin{split}
\underset{\sbQ_i^{(m)}\geq0}{\max}~&\sum_{i=1}^{K}w_ir_i^{(m)}\\
\text{subject to}~&\sum_{i=1}^{K}\text{tr}(\bQ_i^{(m)})\le
\lambda_1P_1+\lambda_2P_2
\end{split}
\end{align}
where the noise covariance at the BS of the dual MAC is
$\lambda_1\bA_1+\lambda_2\bA_2$, and $r_i^{(m)}$ denotes the
achievable rate of the $i$th user. The problem
\eqref{prob:innercapdual} is a convex optimization problem that can
be solved via a standard interior point algorithm. With the optimal
solution of the problem \eqref{prob:innercapdual}, the optimal
solution for the problem \eqref{prob:equalcap} can be obtained by
the MAC-BC transmit covariance matrix transformation. We next
consider the minimization problem
\begin{align}\label{prob:gmin}
\underset{\lambda_1\ge0,\lambda_2\ge0}{\min}~g(\lambda_1,\lambda_2).
\end{align}
Since the function $g(\lambda_1,\lambda_2)$ is not necessarily
differentiable, we can solve the problem \eqref{prob:gmin} via the
subgradient algorithm or ellipsoid algorithm. The subgradient of
the function $g({\lambda}_1,{\lambda}_2)$ can be found in the
following proposition. Refer to Appendix \ref{prov:capsubgra} for
the proof.

\begin{theorem}\label{thm:capsubgra}
The subgradient of the function $g({\lambda}_1,{\lambda}_2)$ at
point $[\bar{\lambda}_1, \bar{\lambda}_2]$ is
$[P_1-\text{tr}(\bar{\bQ}\bA_1) ,P_2-\text{tr}(\bar{\bQ}\bA_2) ]$,
where $\bar{\bQ}$ is the optimal solution of the inner layer
optimization problem \eqref{prob:innercapdual} with
$\lambda_1=\bar{\lambda}_1$ and $\lambda_2=\bar{\lambda}_2$.
\end{theorem}

\begin{Remark}
The Lagrangian function of the problem \eqref{prob:capcomp} can be
written as
\begin{align}\label{Lag:capcomp}
\sum_{i=1}^{K}w_ir_i-\mu_1\big(\text{tr}(\bQ\bA_1)-P_1\big)-\mu_2\big(\text{tr}(\bQ\bA_2)-P_2\big)
\end{align}
while the Lagrangian function of the problem \eqref{prob:equalcap}
can be written as:
\begin{align}\label{Lag:equalcap}
\sum_{i=1}^{K}w_ir_i-\lambda\big(\lambda_1\text{tr}(\bQ\bA_1)+\lambda_2\text{tr}(\bQ\bA_2)-\lambda_1P_1-\lambda_2P_2\big).
\end{align}
By observing \eqref{Lag:capcomp} and \eqref{Lag:equalcap}, we can
say that the two Lagrange functions are identical to each other if
we choose $\mu_1=\lambda\lambda_1$ and $\mu_2=\lambda\lambda_2$.
Thus, the auxiliary variables $\lambda_1$ and $\lambda_2$ can be
viewed as the Lagrange dual variables.
\end{Remark}

Since the function $g(\lambda_1,\lambda_2)$ is a convex function,
the subgradient-based algorithm is guaranteed to converge to its
optimal solution \cite{SBoyd03:Subgrad}. According to Proposition
\ref{thm:upbndachiev}, when the minimum of
$g(\lambda_1,\lambda_2)$ is achieved, the optimal solution of the
problem \eqref{prob:equalcap} is equal to that of the problem
\eqref{prob:capcomp}.

In summary, the problem \eqref{prob:capcomp} is solved through a
two-loop algorithm. By exploiting the general BC-MAC duality, the
inner loop searches the optimal solution of
$g(\lambda_1,\lambda_2)$, while the outer loop solves the
$g(\lambda_1,\lambda_2)$ minimization problem via a
subgradient-based iterative algorithm. The convexity of the
function $g(\lambda_1,\lambda_2)$ guarantees that the global
optimal solution is achieved.

\subsection{Relationship to Minimax Duality}
The capacity region computation problem with a per-antenna power
constraint considered in \cite{Yuwei07:perantconst} is a special
case of the problem \eqref{prob:capcomp}. By choosing $\bA$ to be a
diagonal matrix with all its diagonal elements being zero except the
$j$th diagonal element being 1, the constraint $\text{tr}(\bQ\bA)\le
P$ can be viewed as the power constraint for the $j$th antenna of
the BS. Different from the method discussed in Section
\ref{subsect:mlinearconcap}, the dual MAC problem developed in
\cite{Yuwei07:perantconst} has a minimax form as follows:
\begin{align}\label{prob:yuminimax}
\begin{split}
\underset{\hat{\sbQ}}{\min}~\underset{\sbQ_i^{(\text{mac})}\geq0}{\max}~&\sum_{k=1}^{K}w_kr_k^{(\text{mac})}\\
\text{subject to}~&\sum_{i=1}^{K}\text{tr}(\bQ_i^{(\text{mac})})\le
\text{tr}(\mathbf{\Phi})\\
&\text{tr}(\hat{\bQ}\mathbf{\Phi})\le\text{tr}(\mathbf{\Phi})
\end{split}
\end{align}
where $\mathbf{\Phi}$ is a diagonal matrix,
\[
r_k^{(\text{mac})}:=\log\frac{|\sum_{i=1}^{k}\sbH_i^H\sbQ_i^{(\text{mac})}\sbH_i+\hat{\sbQ}|}{|\sum_{i=1}^{k-1}\sbH_i^H\sbQ_i^{(\text{mac})}\sbH_i+\hat{\sbQ}|}
\]
and the diagonal matrix $\hat{\bQ}$ is the noise covariance matrix
of the dual MAC.  The $j$th diagonal element of $\mathbf{\Phi}$ is
the power constraint for the $j$th antenna of the BS. Since the
noise covariance is also an unknown variable, the existing
high-efficiency algorithm for the MAC problem cannot be applied.
Instead, a new interior point method based algorithm is developed in
\cite{Yuwei07:perantconst} to solve \eqref{prob:yuminimax}.

The two constraints in \eqref{prob:yuminimax} have some
redundancy, and can be further simplified via the following two
observations.
\begin{enumerate}
\item The noise covariance constraint
$\text{tr}(\hat{\bQ}\mathbf{\Phi})\le\text{tr}(\mathbf{\Phi})$
holds with equality when the optimal solution is achieved. \item
Given any positive $\alpha$, if we replace the constraint in
\eqref{prob:yuminimax} with
$\sum_{i=1}^{K}\text{tr}(\bQ_i^{(\text{mac})})\le
\alpha\text{tr}(\mathbf{\Phi})$ and
$\text{tr}(\hat{\bQ}\mathbf{\Phi})\le\alpha\text{tr}(\mathbf{\Phi})$,
the optimal value of the problem \eqref{prob:yuminimax} does not
change.
\end{enumerate}

The observation 1) can be shown by observing that if the constraint
$\text{tr}(\hat{\bQ}\mathbf{\Phi})\le\text{tr}(\mathbf{\Phi})$ is
satisfied with inequality, then the minimization part of the problem
\eqref{prob:yuminimax} does not achieve the optimal solution. The
observation 2) can be proved through the KKT conditions. If we
assume that $\bQ_i^*$ and $\hat{\bQ}^*$ are the optimal transmit
covariance matrix and the noise covariance matrix of the problem
\eqref{prob:yuminimax}, respectively, then it is easy to verify that
$\alpha\bQ_i^*$ and $\alpha\hat{\bQ}^*$ satisfy the KKT conditions
of the problem \eqref{prob:yuminimax} after the constraints
replacement, and the optimal value of the problem
\eqref{prob:yuminimax} does not change.

Based on these two observations, we can combine the two constraints
in \eqref{prob:yuminimax} into one constraint
$\sum_{i=1}^{K}\text{tr}(\bQ_i^{(\text{mac})})\le
\text{tr}(\hat{\bQ}\mathbf{\Phi})$, and thus the problem
\eqref{prob:yuminimax} is equivalent to the following problem:
\begin{align}\label{prob:yuminimax1}
\begin{split}
\underset{\hat{\sbQ}}{\min}\underset{\sbQ_i^{(\text{mac})}\geq0}{\max}~&\sum_{k=1}^{K}w_kr_k^{(\text{mac})}\\
\text{subject
to}~&\sum_{i=1}^{K}\text{tr}(\bQ_i^{(\text{mac})})\le\text{tr}(\hat{\bQ}\mathbf{\Phi}).
\end{split}
\end{align}

In our derivation in Section \ref{subsect:mlinearconcap}, we can
formulate a similar minimax optimization problem of the dual MAC.
Combining \eqref{prob:equalcap}, \eqref{prob:innercapdual} and
\eqref{prob:gmin}, we have
\begin{align}\label{prob:equalcapmac}
\begin{split}
\underset{\lambda_1,\lambda_2}{\min}~\underset{\sbQ_i^{(m)}\geq0}{\max}~&\sum_{i=1}^{K}w_ir_i^{(m)}\\
\text{subject to}~&\sum_{i=1}^{K}\text{tr}(\bQ_i^{(m)})\le
\lambda_1P_1+\lambda_2P_2
\end{split}
\end{align}
where the noise covariance matrix of the dual MAC is
$\lambda_1\bA_1+\lambda_2\bA_2$. In the per-antenna power constraint
scenario, the problem \eqref{prob:equalcapmac} becomes
\begin{align}\label{prob:perantminimax}
\begin{split}
\underset{\sbS}{\min}\underset{\sbQ_i^{(m)}\geq0}{\max}~&\sum_{i=1}^{K}w_ir_i^{(m)}\\
\text{subject to}~&\sum_{i=1}^{K}\text{tr}(\bQ_i^{(m)})\le
\text{tr}(\bS\mathbf{\Phi})
\end{split}
\end{align}
where $\bS=\text{diag}(\lambda_1,\cdots,\lambda_{N_t})$. The problem
\eqref{prob:perantminimax} is identical to the problem
\eqref{prob:yuminimax1} by noting that $\bS=\hat{\bQ}$ and
$\bQ_i^{(m)}=\bQ_i^{(\text{mac})}$. Therefore, the problem
\eqref{prob:yuminimax} and the problem \eqref{prob:perantminimax}
are equivalent to each other.

Although the general BC-MAC duality in Section \ref{sect:duality}
and the minimax duality in \cite{Yuwei06:dualityminmax} have
substantially different formulation, the ways by which they handle
the multiple linear constraints are equivalent essentially. The
general BC-MAC duality based method divides the process into two
steps: dual MAC problem solving, and multiple constraints handling;
while the minimax duality combines the two steps together.
Essentially, only the dual MAC problem solving step exploits the
special BC-MAC reciprocity relationship, and the multiple
constraints handling step is not specific for the BC problem, i.e.,
it can be applied to solve any optimization problem with multiple
constraints. The new method decouples the two steps, and thus has
more flexibility in exploiting the existing algorithm for a MAC.
Moreover, the minimax problem \eqref{prob:yuminimax} is specifically
formulated to solve the per-antenna power constraint problem.

\subsection{Nonlinear Constraint}\label{subsect:nonlinear}

According to the proof of Proposition \ref{thm:SINRduality}, the
general BC-MAC duality requires that the transmit covariance
matrix of the BC be subject to a linear constraint. In this
subsection, we consider the case with a convex but nonlinear
constraint. The capacity computation under a nonlinear constraint
can be formulated as follows:
\begin{align}\label{prob:nonlinear}
\begin{split}
\underset{\sbQ\geq0}{\max}~&\sum_{i=1}^{K}w_ir_i\\
\text{subject to}~&f(\bQ)\le 0
\end{split}
\end{align}
where the function $f(\bQ)$ is a general nonlinear convex
function. Since the general BC-MAC duality does not hold generally
under the nonlinear constraint, the problem \eqref{prob:nonlinear}
cannot be solved directly. However, as shown in the following
proposition, the nonlinear constraint problem
\eqref{prob:nonlinear} is equivalent to a single linear constraint
problem (refer to Appendix \ref{prov:equlinear} for the proof).

\begin{theorem}\label{thm:equlinear}
There always exists a linear constraint problem as follows
\begin{align}\label{prob:equlinear}
\begin{split}
\underset{\sbQ\geq0}{\max}~&\sum_{i=1}^{K}w_ir_i\\
\text{subject to}~&\text{tr}(\bA\bQ_i)\le 0
\end{split}
\end{align}
with $\bA$ denoting a constant matrix such that it has the same
solution as the original problem \eqref{prob:nonlinear}.
\end{theorem}

\begin{Remark}
Proposition \ref{thm:equlinear} illustrates that the nonlinear
non-convex optimization problem \eqref{prob:nonlinear} can be
transformed into an equivalent linear constraint problem
\eqref{prob:equlinear}, which can be solved by making use of the
general BC-MAC duality in Section \ref{sect:duality}. However,
according to the proof of Proposition \ref{thm:equlinear}, the
parameter $\bA$ in problem \eqref{prob:equlinear} cannot be
obtained without the optimal solution $\bQ^*$ of the problem
\eqref{prob:nonlinear}. Note that it is impossible to obtain
$\bQ^*$ before solving this problem, and thus the problem
\eqref{prob:nonlinear} cannot be solved by using its equivalence
with the problem \eqref{prob:equlinear}. In the following, we
present an example to illustrate an iterative algorithm that can
find a set of linear constraints, and these linear constraints can
be used to approximate the original nonlinear constraint.
\end{Remark}

\begin{figure}[tbh]
      \centering
     \includegraphics[width = 120mm]{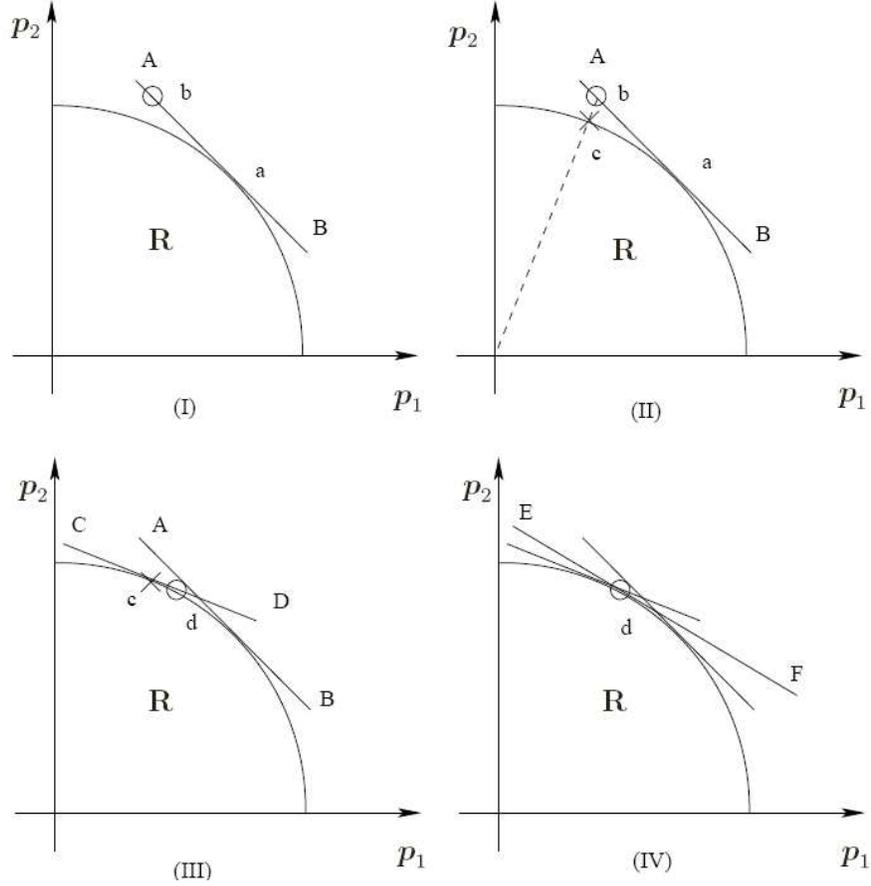}
     \caption{The iterative process for the nonlinear constraint problem. (I) Select an initial point `a', plot a tangent line `AB', and solve the optimization problem with a single linear constraint with respect to `AB', where point `b' corresponds to the optimal solution. (II) Find a point `c', which is closest to `b', on the boundary of the feasible region. (III) Plot a tangent line `CD', and solve the optimization problem with the constraints `AB' and `CD', where `d' is assumed to be the optimal solution. (IV) So on and so forth, until the optimal solution is achieved.}
     \label{fig:nonlinear}
\end{figure}

\begin{example}
For simplicity, we consider the MIMO BC with $K=2$, $N_t=2$, and
$N_r=2$. The transmit covariance matrix is subject to a nonlinear
constraint:
$\big(\text{tr}(\bQ\bA_1)\big)^2+\big(\text{tr}(\bQ\bA_2)\big)^2\le
P$, where $\bA_1=\left [\begin{array}{ccc}1& 0\\0&
0\end{array}\right ]$, and $\bA_2=\left [\begin{array}{ccc}0& 0\\0&
1\end{array}\right ]$. Thus, the capacity computation under the
nonlinear constraint can be formulated as follows:
\begin{align}\label{prob:nonlinear1}
\begin{split}
\underset{\sbQ\geq0}{\max}~&\sum_{i=1}^{2}w_ir_i\\
\text{subject
to}~&\big(\text{tr}(\bQ\bA_1)\big)^2+\big(\text{tr}(\bQ\bA_2)\big)^2\le
P.
\end{split}
\end{align}
As shown in Fig. \ref{fig:nonlinear}, by defining
$p_1:=\text{tr}(\bQ\bA_1)$ and $p_2:=\text{tr}(\bQ\bA_2)$, the
feasible region $\mathbf{R}:~\{p_1,p_2|p_1^2+p_2^2\le P,~p_1\ge 0,
p_2\ge0\}$ is a quarter circle in the nonnegative orthant.
According to Proposition \ref{thm:equlinear}, the problem
\eqref{prob:nonlinear1} is equivalent to the following problem
\begin{align}\label{prob:equlinear1}
\begin{split}
\underset{\sbQ\geq0}{\max}~&\sum_{i=1}^{2}w_ir_i\\
\text{subject to}~&c_1\text{tr}(\bQ\bA_1)+c_2\text{tr}(\bQ\bA_2)\le
c_3
\end{split}
\end{align}
where $c_i=\text{tr}(\bQ^*\bA_i)$, $i=1,2$, $\bQ^*$ is the optimal
solution of the problem \eqref{prob:nonlinear1}, and $c_3=P$.

To present the iterative process, we use a graphical illustration as
shown in Fig. \ref{fig:nonlinear}. Firstly, we arbitrarily select a
point `a' on the boundary of the nonlinear region $\mathbf{R}$, and
draw a tangent line to the $p_1^2+p_2^2= P$ curve through the
selected point. The tangent line corresponds to a linear constraint
$\mathbf{C}^{(1)}:~c_1^{(1)}\text{tr}(\bQ\bA_1)+c_2^{(1)}\text{tr}(\bQ\bA_2)=
c_3^{(1)}$, where the superscript `1' denotes the index of the
linear constraints. The weighted sum rate maximization problem with
the constraint $\mathbf{C}^{(1)}$ can be solved through the general
duality. Assume that $\bQ^{(1)}$ and $r^{(1)}$ are the corresponding
optimal transmit covariance matrix and the obtained optimal weighted
sum rate, respectively. Since the feasible region of the problem
\eqref{prob:nonlinear1} is a subset of the feasible region of
$\mathbf{C}^{(1)}$, $r^{(1)}$ is an upper bound on that of the
original problem \eqref{prob:nonlinear1}. The optimal transmit
covariance matrix $\bQ^{(1)}$, corresponding to the point $b$ in
Fig. \ref{fig:nonlinear}, is not feasible for the original problem.
Secondly, we find the point `c' on the boundary of the region
$\mathbf{R}$ to be closest to the point `b', and draw a new tangent
line through the point `c'. The new tangent line corresponds to a
linear constraint
$\mathbf{C}^{(2)}:~c_1^{(2)}\text{tr}(\bQ\bA_1)+c_2^{(2)}\text{tr}(\bQ\bA_2)=
c_3^{(2)}$. By solving the weighted sum rate maximization problem
with two constraints $\mathbf{C}^{(1)}$ and $\mathbf{C}^{(2)}$, a
new optimal weighted sum rate $r^{(2)}$ is obtained, where $r^{(1)}>
r^{(2)}$. The iterative process will continue until
$\big(\text{tr}(\bQ^{(n)}\bA_1)\big)^2+\big(\text{tr}(\bQ^{(n)}\bA_2)\big)^2\le
P+\epsilon$ holds, where $\bQ^{(n)}$ denotes the optimal solution of
the $n$th iterative step, and $\epsilon$ denotes the prescribed
accuracy requirement. The obtained $r^{(n)}$ forms a non-increasing
sequence with the lower bound on the optimal solution of the problem
\eqref{prob:nonlinear1}. Thus, the algorithm will converge to the
optimal solution.
\end{example}

\section{Beamforming Problem}\label{sect:beamforming}

In the preceding section, the capacity computation for the MIMO BC
is considered. In this section, we consider the MISO BC from a
beamforming perspective\footnote{There are two reasons that we do
not consider the MIMO BC scenario. 1) For the MIMO BC case, due to
the multiple data streams, to impose the SINR constraint for each
data stream is only appropriate when the independent encoding for
each data stream is applied. In the previous work, the beamforming
problems are considered only under the MISO
scenario\cite{Boche04:sinr_bal,Shamai2006:conicoptimazation}. 2)
According to the definition \eqref{def:SINRbc}, the SINR of the BC
is neither convex nor concave with respect to $\bv$ and $\bu$ when
$N_t>1$. Therefore, the MIMO optimization problem with SINR
constraints is not a convex problem, and cannot be solved
efficiently.}. Depending on the objective function and the
constraints, the beamforming optimization problems can be divided
into two classes. One is the SINR balancing problem
\cite{Xu98:sinr_bal,Boche04:sinr_bal}, i.e., maximizing the minimum
SINR among all the users. The other one is the power minimization
problem with SINR constraints
\cite{Boche2005:iterativeDPC,Yuwei07:perantconst,Shamai2006:conicoptimazation},
i.e., minimizing some power function with SINR constraints. In the
case of $N_r=1$, the SINR of the $i$th user under the nonlinear
encoding and decoding scheme can be expressed as follows:
\begin{align}\label{def:sinrnonlinear}
\text{SINR}_i=\frac{|\bh_i^H\bu_i|}{\sum_{j=i+1}^{K}|\bh_i^H\bu_k|^2+1}
\end{align}
where $\bh_i$ denotes the $N_t\times 1$ channel vector from the BS
to the $i$th user, and $\bu_i$ denotes the transmit beamforming
vector. Under the linear encoding and decoding scheme, the
corresponding SINR can be expressed as follows:
\begin{align}\label{def:sinrlinearbeam}
\text{SINR}_i^{(l)}=\frac{|\bh_i^H\bu_i|}{\sum_{j=1,j\neq
i}^{K}|\bh_i^H\bu_k|^2+1}.
\end{align}
It has been shown that the SINR constraint can be transformed into
semidefinite programming (SDP) or SOC form
\cite{Shamai2006:conicoptimazation}, and the corresponding
beamforming problem can be solved via the standard interior point
algorithm. However, the standard algorithm does not exploit the
special structure of the problem, and may be computationally
expensive. A number of efficient iterative algorithms have been
proposed to solve the beamforming problem. In this section, we
will combine these iterative algorithms and the general BC-MAC
duality to solve the beamforming problem with multiple linear
constraints. The methods discussed in this section are applicable
to both SINR definitions \eqref{def:sinrnonlinear} and
\eqref{def:sinrlinearbeam}.

\subsection{SINR Balancing Problem}\label{subsect:sinrbalancing}

In this subsection, we apply the general BC-MAC duality in Section
\ref{sect:duality} to solve the SINR balancing problem with multiple
linear constraints:
\begin{align}\label{prob:bal2}
\begin{split}
\underset{\sbQ\geq0,\alpha\geq0}{\max}~&\alpha\\
\text{subject to}~&\text{SINR}_{i}\ge \alpha
\gamma_{i},~\forall ~i\\
&\text{tr}(\bQ\bA_1)\le P_1\\
&\text{tr}(\bQ\bA_2)\le P_2.
\end{split}
\end{align}
Note that $\bQ=\sum_{i=1}^{K}p_i\bu_i\bu_i^H$ in this case, and
thus the problem can be viewed as a joint beamforming and power
allocation problem. The SINR balancing problem for the BC has been
studied in \cite{Xu98:sinr_bal,Boche04:sinr_bal}. However, due to
the limitations of the conventional BC-MAC duality, previous
results can be applied only to the case in which there is a single
sum power constraint at the BS. Furthermore, the SINR balancing
problem for the MAC with multiple constraints has been studied in
\cite{lan07:jsac}, where it is shown that the multiple constraints
can be completely decoupled. However, the decoupling property does
not hold for the BC scenario.

To solve the problem \eqref{prob:bal2}, we first consider the
following problem:
\begin{align}\label{prob:equalbal2}
\begin{split}
g_{\text{bal}}(\lambda_1,\lambda_2):=&\underset{\bu_{i},p_{i},\alpha}{\max}~\alpha\\
\text{subject to}~&\text{SINR}_{i}\ge \alpha
\gamma_{i},~\forall ~i\\
&\lambda_1\text{tr}(\bQ\bA_1)+\lambda_2\text{tr}(\bQ\bA_2)\le
\lambda_1P_1+\lambda_2 P_2
\end{split}
\end{align}
where $\lambda_1$ and $\lambda_2$ are auxiliary variables. Similar
to Proposition \ref{thm:upperbnd}, we can prove that the optimal
solution of the problem \eqref{prob:equalbal2} is an upper bound on
that of the problem \eqref{prob:bal2}. The problem
\eqref{prob:equalbal2} can be transformed into its dual MAC problem
via the general BC-MAC duality and efficiently solved by the
iterative algorithm in \cite{Boche04:sinr_bal}. Moreover, the
minimization problem
$\min_{\lambda_1,\lambda_2}g_{\text{bal}}(\lambda_1,\lambda_2)$ can
be solved through the subgradient algorithm or ellipsoid algorithm.
The convexity of the function $g_{\text{bal}}(\lambda_1,\lambda_2)$
guarantees the convergence of the subgradient-based algorithm.
Similar to the capacity region computation problem in Section
\ref{subsect:mlinearconcap}, it can be proved that the algorithm
converges to an optimal solution of the problem \eqref{prob:bal2}.

\begin{Remark}
The method to process multiple linear constraints in the SINR
balancing problem is identical to that in the capacity computation
problem. We first introduce an upper bound function, the solution of
which can be obtained via the general BC-MAC duality. We next
compute the minimum value of the upper bound function via a
subgradient-based algorithm. Note that the iterative algorithm in
\cite{Boche04:sinr_bal} cannot be applied to the minimax duality
case as the iterative algorithm requires the explicit expression of
the noise covariance matrix. From this perspective, the general
BC-MAC duality has broader application than the minimax duality.
\end{Remark}

\subsection{Power Balancing Problem}\label{subsect:powermin}

In this subsection, we consider the power balancing problem with
SINR constraints. Mathematically, the problem is formulated as
follows:
\begin{align}\label{prob:minpower}
\begin{split}
\underset{\sbQ\geq0,\alpha\geq0}{\min}~&\alpha\\
\text{subject to}~&\text{SINR}_{i}\ge \gamma_{i}\\
&\text{tr}(\bQ\bA_1)\le \alpha P_1\\
&\text{tr}(\bQ\bA_2)\le \alpha P_2.
\end{split}
\end{align}
Since the problem \eqref{prob:minpower} has multiple power
constraints, the general BC-MAC duality cannot be applied
directly. By introducing two auxiliary variables $\lambda_1$ and
$\lambda_2$, we transform the problem \eqref{prob:minpower} into
the following single power constraint problem:
\begin{align}\label{prob:innerminpower}
\begin{split}
g_{\text{pow}}(\lambda_1,\lambda_2):=&\underset{\sbQ\geq0,\alpha\geq0}{\min}~\alpha\\
\text{subject to}~&\text{SINR}_{i}\ge \gamma_{i}\\
&\lambda_1\text{tr}(\bQ\bA_1)+\lambda_2\text{tr}(\bQ\bA_2)\le\alpha(\lambda_1
P_1+\lambda_2P_2).
\end{split}
\end{align}
Similar to Proposition \ref{thm:upperbnd}, the optimal solution of
the problem \eqref{prob:innerminpower} is a lower bound on that of
the problem \eqref{prob:minpower}. Thanks to the general BC-MAC
duality, the MIMO BC problem \eqref{prob:innerminpower} is
equivalent to its dual MAC problem as follows:
\begin{align}\label{prob:innerminpmac}
\begin{split}
\underset{\sbQ\geq0}{\min}~&\sum_{i=1}^{K}\text{tr}(\bQ_i^{(m)})\\
\text{subject to}~&\text{SINR}_{i}^{(m)}\ge \gamma_{i}
\end{split}
\end{align}
where the noise covariance of this dual MAC is
$\lambda_1\bA_1+\lambda_2\bA_2$. By solving the problem
\eqref{prob:innerminpmac} with the algorithm in
\cite{Boche2005:iterativeDPC}, and utilizing the MAC-BC transmit
covariance matrix transformation, the optimal solution of the
problem \eqref{prob:innerminpower} can be obtained. Next, we
consider a maximization problem as follows:
\begin{align}\label{prob:outerminpower}
\underset{\lambda_1\ge0,\lambda_2\ge0}{\max}g_{\text{pow}}(\lambda_1,\lambda_2).
\end{align}
Similar to Proposition \ref{thm:capsubgra}, we have the following
result concerning the subgradient of the function
$g_{\text{pow}}(\lambda_1,\lambda_2)$.
\begin{theorem}\label{thm:minpsubg}
The subgradient of the function
$g_{\text{pow}}(\lambda_1,\lambda_2)$ at
$[\tilde{\lambda}_1,\tilde{\lambda}_2]$ is
$[\text{tr}(\tilde{\bQ}\bA_1)- P_1,\text{tr}(\tilde{\bQ}\bA_2)-
P_2]$, where $\tilde{\bQ}$ is the optimal solution of the inner
layer problem \eqref{prob:innerminpower} with
$\lambda_1=\tilde{\lambda}_1$ and $\lambda_2=\tilde{\lambda}_2$.
\end{theorem}

The proof of Proposition \ref{thm:minpsubg} is similar to that of
Proposition \ref{thm:capsubgra}, and thus is omitted here. With
Proposition \ref{thm:minpsubg}, the maximization problem
\eqref{prob:outerminpower} can be solved through the subgradient
algorithm or the ellipsoid algorithm. Similar to Proposition
\ref{thm:upbndachiev}, when the maximum of
$g_{\text{pow}}(\lambda_1,\lambda_2)$ is achieved, the optimal
solution of the problem \eqref{prob:innerminpower} is equal to that
of the problem \eqref{prob:minpower}

\begin{Remark}
The maximum per-antenna power constraint minimization problem was
considered in \cite{Yuwei07:perantconst}, which is a special case of
the problem \eqref{prob:minpower}. In \cite{Yuwei07:perantconst},
the problem is transformed into its minimax dual MAC problem, in
which the noise covariance matrix of its dual MAC is an unknown
variable. A subgradient-based iterative algorithm is developed
therein to obtain its optimal noise covariance. However, since the
noise covariance matrix appears in both the constraints and the
objective function, it is difficult to have a routine method to
obtain its subgradient in \cite{Yuwei07:perantconst}. While, in
contrast, due to the clear physical meaning of the variables
$\lambda_i$ (Lagrange dual variables with respect to some
constraints), the subgradient of the lower bound function
$g_{\text{pow}}(\lambda_1,\lambda_2)$ can be readily obtained.
\end{Remark}

\section{Numerical Results}\label{sect:simulation}

In this section, we present several numerical results to illustrate
the effectiveness of the proposed algorithms. For simplicity, we
consider a MIMO BC with $K=2$, $N_t=2$, $N_r=2$ for the capacity
computation problem and with $K=2$, $N_t=2$, $N_r=1$ for the
beamforming problem. The noise covariance matrix at each user is
assumed to be an identity matrix.

\subsection{Capacity Region of the MIMO BC}

\begin{figure}[tbh]
      \centering
     \includegraphics[width = 120mm]{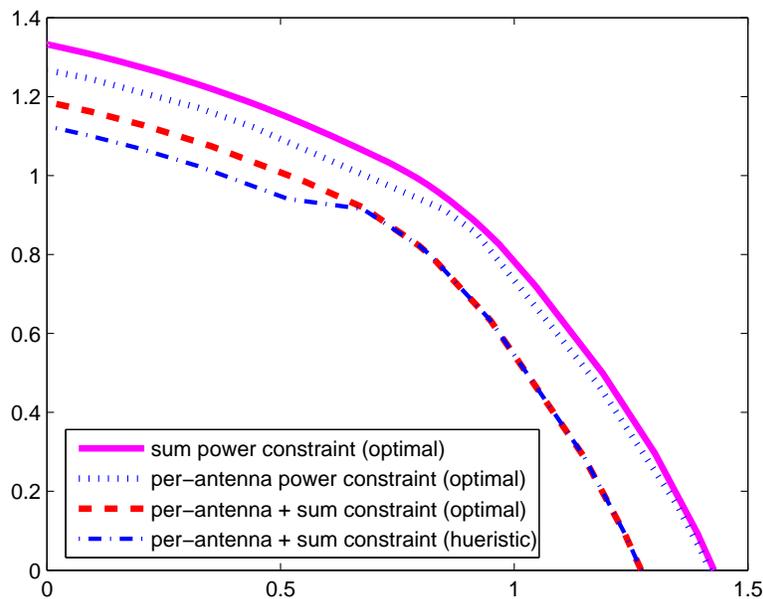}
     \caption{The capacity regions for the MIMO BC with various power constraints.}
     \label{fig:capacity}
\end{figure}

In this example, we compute the capacity regions of the MIMO BC with
a sum power constraint, a per-antenna power constraint and sum power
plus per-antenna power constraints, separately. The sum power
constraint is taken to be 10, and the per-antenna power is taken to
be 5. For the sum power plus per-antenna power constraint case, the
sum power constraint is 8, and the per-antenna power constraint is
5. The channel matrices are chosen to be $\bH_1=\left
[\begin{array}{ccc}1& 0\\0.2& 0.6\end{array}\right ]$ and
$\bH_2=\left [\begin{array}{ccc}0.5 &0\\ 0.2 &1\end{array}\right ]$.
For the case with a sum power constraint, the algorithm is similar
to that in \cite{rzhang06:powerregion}. For the case with a
per-antenna power constraint the subgradient-based iterative
algorithm developed in Section \ref{sect:capacity} is applied. For
the sum power plus per-antenna power constraints case, two different
algorithms are adopted. The first one is the subgradient-based
iterative algorithm. The second algorithm is a heuristic algorithm,
and is based on the result obtained in the case with a sum power
constraint. With the sum power constraint solution, the transmit
covariance matrix is normalized such that each antenna's power
satisfies the per-antenna power constraint. The regions obtained by
these algorithms are shown in Fig. \ref{fig:capacity}. Since the
heuristic algorithm obtains the suboptimal solution, the fourth line
is just an achievable rate region of the MIMO BC with sum power plus
per-antenna power constraints. Moreover, since the per-antenna power
constraint is stricter than the sum power constraint, the capacity
region of the case with a sum power constraint is larger than that
of the case with a per-antenna power constraint.

\subsection{Weighted Sum Rate Maximization With Nonlinear Constraint}

\begin{figure}[htb]
\begin{minipage}[b]{1.0\linewidth}
  \centering
  \includegraphics[width = 120mm]{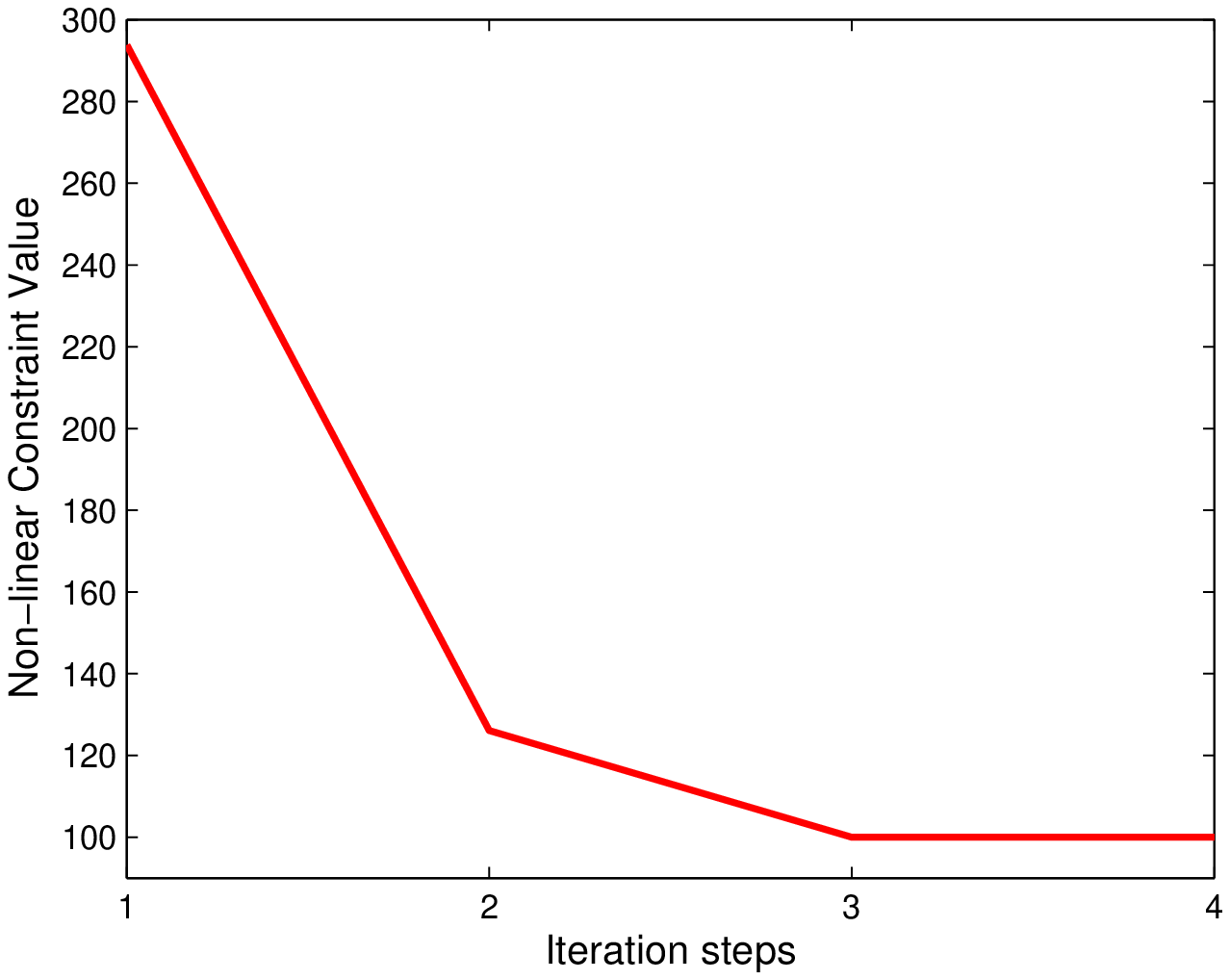}
  \centerline{(a) The value of $\big(\text{tr}(\bQ\bA_1)\big)^2+\big(\text{tr}(\bQ\bA_2)\big)^2$ versus the iteration steps.}\medskip
\end{minipage}
\begin{minipage}[b]{1.0\linewidth}
  \centering
  \includegraphics[width = 120mm]{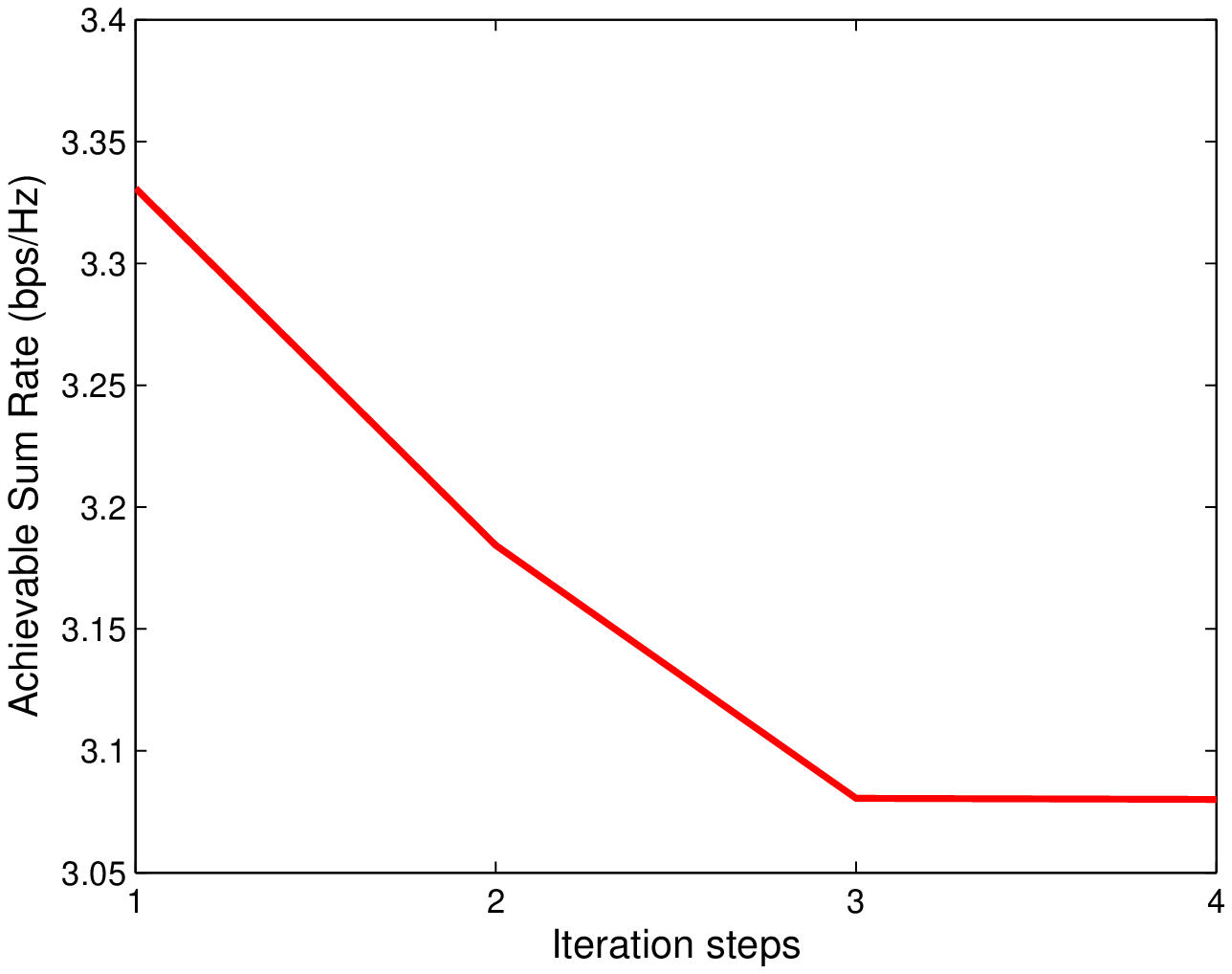}
  \centerline{(b) Achieved sum rate versus the iteration steps.}\medskip
\end{minipage}
\caption{ The convergence behavior of the subgradient-based
algorithm for the weighted sum rate maximization problem with a
non-linear constraint. } \label{fig:nonlinear1}
\end{figure}

This subsection is to present the simulation result of the Example
1 in Section \ref{subsect:nonlinear}. Suppose that $P$ in
\eqref{prob:nonlinear} is 100. The channel matrices are chosen to
be $\bH_1=\left [\begin{array}{ccc}2& 0\\0.5& 0.6\end{array}\right
]$ and $\bH_2=\left [\begin{array}{ccc}0.3& 0.2\\ 0&
1.5\end{array}\right ]$. In Fig. \ref{fig:nonlinear1} (a), the
non-linear constraint function values are plotted versus the
iteration steps. It can be observed that the non-linear transmit
covariance constraint is satisfied when the optimal solution is
achieved. In Fig. \ref{fig:nonlinear} (b), the achieved sum rates
are plotted versus the iteration steps. The curve in Fig.
\ref{fig:nonlinear1} (b) is non-increasing, since the results in
the former steps are obtained by solving the weighted sum rate
problem with relaxed constraints.

\subsection{SINR Balancing With Multiple Linear Transmit Covariance Constraints}

\begin{figure}[htb]
\begin{minipage}[b]{1.0\linewidth}
  \centering
  \includegraphics[width = 120mm]{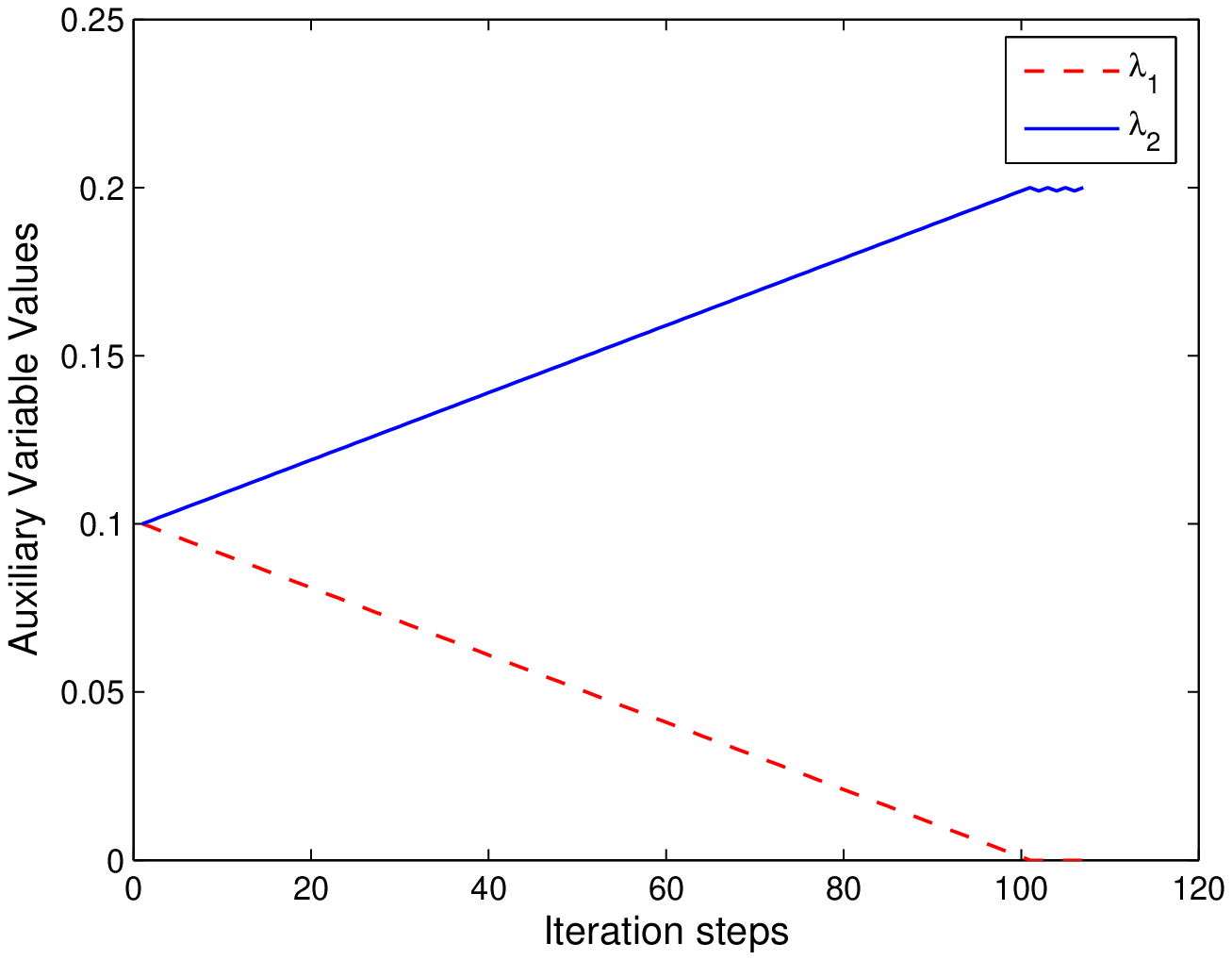}
  \centerline{(a) Auxiliary variable values versus the iteration steps.}\medskip
\end{minipage}
\begin{minipage}[b]{1.0\linewidth}
  \centering
  \includegraphics[width = 120mm]{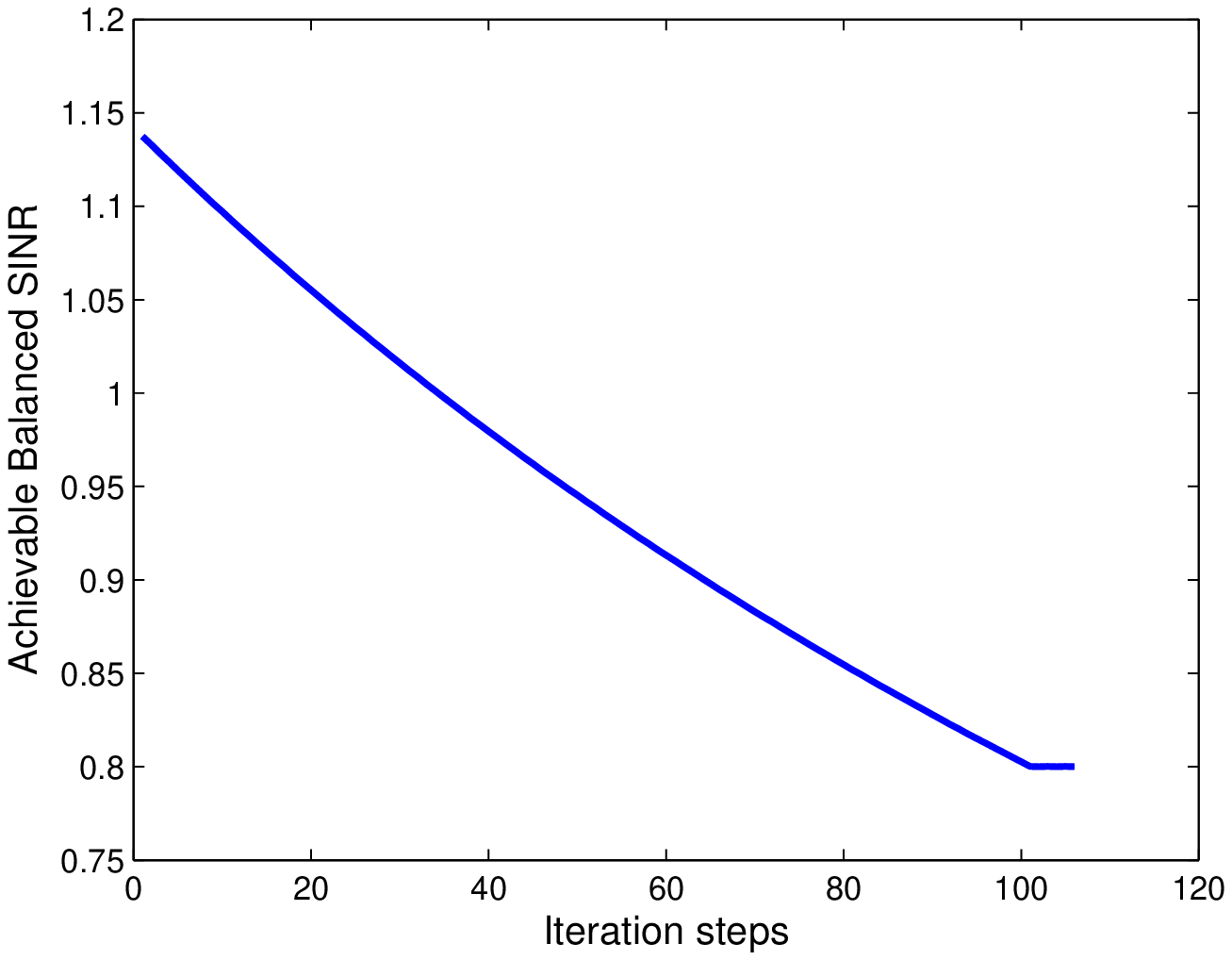}
  \centerline{(b) Achieved balanced SINR versus the iteration steps.}\medskip
\end{minipage}
\caption{ The convergence behavior of the subgradient-based
algorithm for the SINR balancing problem with per-antenna power
constraint. } \label{fig:SINR}
\end{figure}

In this example, we consider the SINR balancing problem with a
per-antenna power constraint. We assume that each antenna's
transmit power is subject to the constraint 5, and each user's
target SINR is $\gamma_i=1$, for $i=1,2$. The channel matrix is
chosen to be $\bH_1=\left [\begin{array}{ccc}1 &0\\0.5
&0.6\end{array}\right ]$ and $\bH_2=\left [\begin{array}{ccc}0.4
&0\\ 0.5& 1.5\end{array}\right ]$. The convergence behavior of the
algorithm in Section \ref{subsect:sinrbalancing} is shown in Fig.
\ref{fig:SINR}. The achieved SINR for each iteration is plotted in
Fig. \ref{fig:SINR} (a). It can be observed that the curve in Fig.
\ref{fig:SINR} (a) is non-increasing, and the achieved SINR for
each iteration is greater than or equal to the final result. This
is because that the optimal solution of the problem
\eqref{prob:equalbal2} is an upper bound on that of the original
problem \eqref{prob:bal2}. The auxiliary variable values are
plotted in Fig. \ref{fig:SINR} (b). It can be seen from this
figure that only one constraint is active when the optimal
solution is achieved.

\section{Conclusions}\label{sect:conclusions}

In this paper, we have established a general Gaussian BC-MAC
duality, where the BC is subject to a general linear constraint
and the MAC is subject to a weighted sum power constraint. This
general BC-MAC duality can be applied to solve the capacity
computation and beamforming optimization problems with multiple
convex linear/nonlinear constraints. The relationship between the
new method and the previous minimax-duality based method has also
been discussed. Moreover, it has been shown that, compared to the
minimax duality the general BC-MAC duality offer greater
flexibility for solving BC optimization problems. This new duality
also generalizes the conventional Gaussian BC-MAC duality.

\def\appref#1{Appendix~\ref{#1}}
\appendix
\renewcommand{\thesubsection}{\Alph{subsection}}
\makeatletter
\renewcommand{\subsection}{%
\@startsection {subsection}{2}{\z@ }{2.0ex plus .5ex minus .2ex}%
{-1.0ex plus .2ex}{\it }} \makeatother

\subsection{Proof of Corollary \ref{cor:rateregion}}\label{prov:cor}

The corollary can be derived from Proposition
\ref{thm:SINRduality} and the relationship between the SINR and
the achievable rate \eqref{eq:rateSINR}. We will verify the
corollary directly via covariance matrix transformation, as
follows.

The achievable rate of the $i$th user of the dual MAC can be written
as
\begin{align}
r_i^{(m)}&=\log\frac{|\sbA+\sum_{k=1}^{i}\sbH_k^H\sbQ_{k}^{(m)}\sbH_k|}{|\sbA+\sum_{k=1}^{i-1}\sbH_k^H\sbQ_{k}^{(m)}\sbH_k|}\\
&=\log\frac{|\sbI+\sum_{k=1}^{i}\sbA^{-1}\sbH_k^H\sbQ_{k}^{(m)}\sbH_k|}{|\sbI+\sum_{k=1}^{i-1}\sbA^{-1}\sbH_k^H\sbQ_{k}^{(m)}\sbH_k|}\\
&=\log\frac{|\sbI+\sum_{k=1}^{i}(\sbH_k\sbA^{-1/2})^H\sbQ_{k}^{(m)}\sbH_k\sbA^{-1/2}|}{|\sbI+\sum_{k=1}^{i-1}(\sbH_k\sbA^{-1/2})^H\sbQ_{k}^{(m)}\sbH_k\sbA^{-1/2}|}\label{eq:equMAC}
\end{align}
where the eigenvalue decomposition of $\bA^{-1}$ is
$\bU^H\mathbf{\Lambda}\bU$, and
$\bA^{-1/2}=\bU^H\mathbf{\Lambda}^{1/2}$. According to
\eqref{eq:equMAC}, the MAC can be viewed as a \emph{virtual MAC}
with $(\bH_k\bA^{-1/2})^H$ being its channel matrix and its noise
covariance matrix being an identity matrix. By exploiting the BC-MAC
covariance algorithm in \cite{Jindal03:sumcapacity}, the achievable
region of the virtual MAC is equal to the \emph{virtual BC} with
$\bH_k\bA^{-1/2}$ being its channel matrix. Thus, the achievable
rate of the $i$ user of the virtual BC can be written as
\begin{align}
r_i=\log\frac{|\sbI+\sum_{k=i}^{K}\sbH_k\sbA^{-1/2}\sbQ_{k}(\sbH_k\sbA^{-1/2})^H|}{|\sbI+\sum_{k=i+1}^{K}\sbH_k\sbA^{-1/2}\sbQ_{k}(\sbH_k\sbA^{-1/2})^H|}
\end{align}
where $\sum_{k=1}^{K}\text{tr}(\bQ_k)=P$. By defining
$\bQ^{(b)}_k=\bA^{-1/2}\bQ_{k}(\bA^{-1/2})^H$, we have
$\sum_{k=1}^{K}\text{tr}(\bA\bQ^{(b)}_k)=P$. Thus, the achievable
rate of the dual MAC with sum power constraint $P$ is also
achievable for the primal BC with the constraint
$\text{tr}(\bA\bQ)\leq P$. Similarly, we can prove that the
achievable rate of the primal BC with the constraint
$\text{tr}(\bA\bQ)\leq P$ is also achievable for the dual MAC with
the sum power constraint $P$. The proof follows.
$~~~~~~~~~~~~~~~~~~~~~~~~~~~~~~~~~~~~~~~~~~~~~~~~~~~~~~~~~~~~~~~~~~~~~~~~~~~~~~~~~~~~~~~~~~~~~\blacksquare$
\subsection{Proof of Proposition
\ref{thm:capsubgra}}\label{prov:capsubgra}

According to the definition of the subgradient, if $[s_1, s_2]$ is
the subgradient of $g(\lambda_1, \lambda_2)$ at point
$[\bar{\lambda}_1, \bar{\lambda}_2]$, then we have
$g(\tilde{\lambda}_1,\tilde{\lambda}_2)\ge
g(\bar{\lambda}_1,\bar{\lambda}_2)+[s_1,
s_2]\cdot\big([\tilde{\lambda}_1,
\tilde{\lambda}_2]-[\bar{\lambda}_1, \bar{\lambda}_2]\big)^H$ for
any $[\tilde{\lambda}_1, \tilde{\lambda}_2]$.

The Lagrange function of the problem \eqref{prob:equalcap} can be
written as
\begin{align}
L(\bQ,\lambda)=\sum_{i=1}^Kw_ir_i-\lambda\Big(\lambda_1\text{tr}(\bQ\bA_1)&+\lambda_2\text{tr}(\bQ\bA_2)-
\lambda_1P_1-\lambda_2P_2\Big).
\end{align}
Thus, the corresponding dual problem is
\begin{align}
\underset{\lambda\ge0}{\min}~\underset{\sbQ}{\max}L(\bQ,\lambda,\lambda_1,\lambda_2).
\end{align}

We have
\begin{align}
&g(\tilde{\lambda}_1,\tilde{\lambda}_2)-g(\bar{\lambda}_1,\bar{\lambda}_2)\\
=&L(\bQ,\lambda)\big
|_{\sbQ=\tilde{\sbQ},\lambda=\tilde{\lambda}}^{\tilde{\lambda}_1,\tilde{\lambda}_2}-L(\bQ,\lambda)\big
|_{\sbQ=\tilde{\sbQ},\lambda=\bar{\lambda}}^{\bar{\lambda}_1,\bar{\lambda}_2}\\
\ge&L(\bQ,\lambda)\big
|_{\sbQ=\bar{\sbQ},\lambda=\tilde{\lambda}}^{\tilde{\lambda}_1,\tilde{\lambda}_2}-L(\bQ,\lambda)\big
|_{\sbQ=\tilde{\sbQ},\lambda=\bar{\lambda}}^{\bar{\lambda}_1,\bar{\lambda}_2}\\
=&-\tilde{\lambda}\Big(\tilde{\lambda}_1(\text{tr}(\bar{\bQ}\bA_1)-P_1)+\tilde{\lambda}_2(\text{tr}(\bar{\bQ}\bA_2)-P_2)\Big)+\bar{\lambda}\Big(\bar{\lambda}_1(\text{tr}(\bar{\bQ}\bA_1)-P_1)+\bar{\lambda}_2(\text{tr}(\bar{\bQ}\bA_2)-P_2)\Big)\\
=&(\text{tr}(\bar{\bQ}\bA_1)-P_1)(-\tilde{\lambda}\tilde{\lambda}_1+\bar{\lambda}\bar{\lambda}_1)+(\text{tr}(\bar{\bQ}\bA_2)-P_2)(-\tilde{\lambda}\tilde{\lambda}_2+\bar{\lambda}\bar{\lambda}_2)\\
=&(\text{tr}(\bar{\bQ}\bA_1)-P_1)(-\tilde{\lambda}\bar{\lambda}_1+\tilde{\lambda}\bar{\lambda}_1-\tilde{\lambda}\tilde{\lambda}_1+\bar{\lambda}\bar{\lambda}_1)+(\text{tr}(\bar{\bQ}\bA_2)-P_2)(-\tilde{\lambda}\bar{\lambda}_2+\tilde{\lambda}\bar{\lambda}_2-\tilde{\lambda}\tilde{\lambda}_2+\bar{\lambda}\bar{\lambda}_2)\\
=&(\text{tr}(\bar{\bQ}\bA_1)-P_1)(-\tilde{\lambda}\tilde{\lambda}_1+\tilde{\lambda}\bar{\lambda}_1)+(\text{tr}(\bar{\bQ}\bA_1)-P_1)(\bar{\lambda}\bar{\lambda}_1-\tilde{\lambda}\bar{\lambda}_1)\notag\\
&+(\text{tr}(\bar{\bQ}\bA_2)-P_2)(-\tilde{\lambda}\tilde{\lambda}_2+\tilde{\lambda}\bar{\lambda}_2)+(\text{tr}(\bar{\bQ}\bA_2)-P_2)(\bar{\lambda}\bar{\lambda}_2-\tilde{\lambda}\bar{\lambda}_2)\\
=&\tilde{\lambda}[P_1-\text{tr}(\bar{\bQ}\bA_1),~P_2-\text{tr}(\bar{\bQ}\bA_2)]\cdot[\tilde{\lambda}_1-\bar{\lambda}_1,~\tilde{\lambda}_2-\bar{\lambda}_2]^H\label{eq:prov1}
\end{align}
where \eqref{eq:prov1} is due to
$(\text{tr}(\bar{\bQ}\bA_1)-P_1)\bar{\lambda}_1+(\text{tr}(\bar{\bQ}\bA_2)-P_2)\bar{\lambda}_2=0$.
The proof follows. $~~~~~~~~~~~~~~~~~~~~~~~~\blacksquare$

\subsection{Proof of Proposition
\ref{thm:equlinear}}\label{prov:equlinear}

Suppose that $\bQ^*$ is the optimal solution of the problem
\eqref{prob:nonlinear}. The KKT conditions of the problem
\eqref{prob:nonlinear} can be written as
\begin{align}
&\frac{\partial\sum_{i=1}^{K}w_ir_i}{\partial
\sbQ_i}\Big |_{\sbQ_i=\sbQ_i^*}=\lambda f'(\bQ_i)\Big |_{\sbQ_i=\sbQ_i^*}+\mathbf{\Psi}_i,~\forall i\\
&\lambda f(\bQ^*)=0.
\end{align}
Note that $\bQ=\sum_{i=1}^{K}\bQ_i$, and thus we have
$f'(\bQ_i)=f'(\bQ)\frac{\partial \sbQ}{\partial \sbQ_i}=f'(\bQ)$.
Now, let us consider the linear constraint problem
\begin{align}\label{prob:eqlinearpro}
\begin{split}
\underset{\sbQ}{\max}~&\sum_{i=1}^{K}w_ir_i\\
\text{subject to}~&\text{tr}(\bA\bQ)\le 0
\end{split}
\end{align}
where $\bA=f'(\bQ)\Big |_{\sbQ=\sbQ^*}$. The KKT conditions for
the problem \eqref{prob:eqlinearpro} are
\begin{align}
&\frac{\partial\sum_{i=1}^{K}w_ir_i}{\partial
\sbQ_i}\Big |_{\sbQ_i=\sbQ_i^*}=\lambda f'(\bQ_i)\Big |_{\sbQ_i=\sbQ_i^*}+\mathbf{\Psi}_i,~\forall i\\
&\lambda f(\bQ^*)=0.
\end{align}

It is easy to observe that $\bQ^*$ satisfies the KKT conditions of
the problem \eqref{prob:eqlinearpro}. Combining this with
Proposition \ref{thm:kktsuf}, we can conclude that $\bQ^*$ is the
optimal solution of problem \eqref{prob:eqlinearpro}. The proof
follows.$\blacksquare$

{\linespread{1.5}

}
\end{document}